 \documentclass[final,authoryear,5p,times,twocolumn]{elsarticle}

\usepackage{graphicx}
\graphicspath{{figures/}}

\usepackage{amssymb}

\usepackage{amsmath}

\usepackage{supertabular}

\journal{Icarus}

\usepackage{wasysym}

\usepackage{natbib}
\bibpunct{(}{)}{;}{a}{}{,}
\usepackage{pstricks}

\usepackage{hyperref}
\hypersetup{
    bookmarks=true,         
    unicode=false,          
    pdftoolbar=true,        
    pdfmenubar=true,        
    pdffitwindow=false,     
    pdftitle={Spectral properties of near-Earth and Mars-crossing asteroids using Sloan photometry}, 
    pdfsubject={Planetary Science}, 
    pdfkeywords={},         
    pdfnewwindow=true,      
    colorlinks=true,        
    linkcolor=gray,         
    citecolor=blue,         
    filecolor=gray,         
    urlcolor=gray           
}

\usepackage{multirow}

\renewcommand{\u}{u$^\prime$}
\newcommand{\g}{g$^\prime$}
\renewcommand{\r}{r$^\prime$}
\renewcommand{\i}{i$^\prime$}
\newcommand{\z}{z$^\prime$}

\newcommand{\eg}{e.g.}
\newcommand{\ie}{i.e.}
\newcommand{\degr}{\ensuremath{^{\textrm{o}}}}
\newcommand{\arcsec}{\ensuremath{^{\prime\prime}}}

\newcommand{\numb}[1]{\textbf{#1}}
\renewcommand{\numb}[1]{#1}

\newcommand{\add}[1]{\textbf{#1}}
\newcommand{\rem}[1]{\textcolor{red}{\sout{#1}}}

\renewcommand{\add}[1]{#1}
\renewcommand{\rem}[1]{}

\begin{document}

\begin{frontmatter}



\title{Spectral properties of near-Earth and Mars-crossing asteroids using Sloan photometry}


\author[imcce,esac]{B. Carry}
  \ead{benoit.carry@oca.eu}
\author[laeff,svo]{E. Solano}
\author[imcce]{S. Eggl}
\author[mit,harvard]{F. E. DeMeo}

\address[imcce]{IMCCE, Observatoire de Paris, PSL Research University, CNRS, Sorbonne Uni-
versit ́es, UPMC Univ Paris 06, Univ. Lille}
\address[esac]{European Space Astronomy Centre, ESA, P.O. Box 78,
  28691 Villanueva de la Ca{\~n}ada, Madrid, Spain}
\address[laeff]{Centro de Astrobiologia (INTA-CSIC), Departamento de
  Astrofisica. P.O. Box 78, E-28691 Villanueva de la Ca{\~n}ada, Madrid, Spain }
\address[svo]{Spanish Virtual Observatory}
\address[mit]{Department of Earth, Atmospheric and Planetary Sciences,
  MIT, 77 Massachusetts Avenue, Cambridge, MA, 02139, USA}
\address[harvard]{Harvard-Smithsonian Center for Astrophysics, 60
  Garden Street, MS-16, Cambridge, MA, 02138, USA}

\begin{abstract}
  The nature and origin of the asteroids orbiting in
  near-Earth space, including those on a potentially hazardous trajectory,
  is of both scientific interest and practical importance. 
  We aim here at determining the taxonomy of a large sample of near-Earth and
  Mars-crosser asteroids and analyze the distribution of these classes with
  orbit. We use this distribution to identify the source regions of
  near-Earth objects and to study the strength of planetary encounters to 
  refresh asteroid surfaces.
  We measure the photometry of these asteroids over four filters
  at visible wavelengths on images taken by the Sloan
  Digital Sky Survey (SDSS).
  These colors are used to classify the asteroids
  into a taxonomy consistent with the widely used Bus-DeMeo taxonomy
  (DeMeo et al., Icarus 202, 2009) based on visible and near-infrared spectroscopy.
  We report here on the taxonomic classification of \numb{206} near-Earth 
  and \numb{776} Mars-crosser asteroids determined from SDSS
  photometry, representing an increase of \numb{40}\% and \numb{663}\% of
  known taxonomy 
  classifications in these populations.
  Using the source region mapper by Greenstreet et al. (Icarus, 217, 2012), we compare
  for the first time the taxonomic distribution among near-Earth and 
  main-belt asteroids of similar diameters. 
  Both distributions agree at \add{the few percent} level for the inner part of
  the Main Belt and we confirm this region as a main source of near-Earth objects.
  The effect of planetary encounters on asteroid surfaces are also studied by
  developing a simple model of forces acting on a surface grain during
  planetary encounter, which provides the minimum distance at which a
  close approach should occur to trigger resurfacing events. 
  By integrating numerically the orbit of the
  \numb{519} S-type and \numb{46} Q-type asteroids in our sample back in time
  for 500,000 years and monitoring their encounter distance with 
  Venus, Earth, Mars, and Jupiter, we seek to understand the
  conditions for resurfacing events. The 
  population of Q-type \add{is} found to present statistically more
  encounters with Venus and the Earth than S-types, \add{although both
  S- and Q-types present the same amount of encounters with Mars}.
\end{abstract}

\begin{keyword}
Near-Earth objects
\sep
Asteroids, composition
\sep
Photometry
\end{keyword}

\end{frontmatter}



\section{Introduction}
  \indent Asteroids are the leftovers of
  the building blocks that accreted to form the planets in the early
  Solar System.
  They are also the progenitors of the constant influx of
  meteorites falling on the planets, including the Earth.
  Apart from the tiny sample of rock from asteroid
  (25\,143) Itokawa brought back by the Hayabusa spacecraft
  \citep{2011-Science-333-Nakamura}, 
  these meteorites represent
  our sole possibility to study in details the composition of asteroids.
  Identifying their source regions is crucial to determine
  the physical conditions and abundances in elements that reigned in
  the protoplanetary nebula around the young Sun \cite[see,
    \eg,][]{2006-MESS2-McSween}. 
  From the analysis of a bolide trajectory, it is possible to
  reconstruct its heliocentric orbit and to find its parent body
  \citep[\eg,][]{2006-MPS-41-Gounelle}, 
  but such determinations have been limited to a few
  objects only \citep{2012-AA-541-Rudawska}. \\ 
  \indent Among the different dynamical classes of asteroids, 
  the near-Earth and Mars-crosser asteroids (NEAs and MCs), whose 
  orbits cross that of the telluric planets,
  form a transient population.
  Their typical lifetime is of a few million years only
  \citep{2002-Icarus-156-Bottke, 2002-AsteroidsIII-4-Morbidelli}
  before being ejected from the Solar System, falling into the Sun, or
  impacting a planet. 
  These populations are therefore constantly replenished by asteroids
  from the main asteroid belt, the largest reservoir of known small
  bodies, between Mars and Jupiter.\\
  \indent The resonances between the orbits of asteroids and that of
  Jupiter have been long thought \citep{1979-Icarus-37-Wetherill,
    1983-Icarus-56-Wisdom} 
  to provide the kick in eccentricity necessary to place
  asteroids on planet-crossing orbits.
  It was later found that the 
  secular resonance $\nu_6$, delimiting the inner edge of the main belt,
  and the 3:1 mean-motion resonance \add{(MMR)} with Jupiter, separating the inner from the
  middle belt, were the most effective, compared to the 5:2
  resonance, for instance, which tends to eject asteroids from the solar system
  \citep[see][for a review]{2002-AsteroidsIII-4-Morbidelli}. 
  The major role played by the $\nu_6$ resonance was confirmed by the 
  comparison between the reflectance spectra of the most common
  meteorites, the ordinary chondrites (OCs, 80\% of all meteorite falls), the
  dominant class in the near-Earth space, the S-type asteroids
  \citep[about 65\% of the observed population,][]{2004-Icarus-170-Binzel},
    and the dynamical family of S-types asteroids linked with (8) Flora in the
    inner belt \citep{2008-Nature-454-Vernazza}. \\
  \indent The NEAs also represent ideal targets for space exploration
  owing to their close distance from Earth.
  This proximity is quantified by the energy required to set a
  spacecraft on a rendezvous trajectory and is often expressed as 
  $\Delta v$ (in km/s), the required change in speed.
  This is the reason why the first mission to an asteroid targeted the
  Amor (433) Eros 
  \citep{2000-Science-289-Veverka}, 
  why all the targets of sample-return missions were selected
  among NEAs: 
  (25\,143) Itokawa for JAXA Hayabusa \citep{2006-Science-312-Fujiwara},
  (101\,955) Bennu for NASA OSIRIS-REx
  \citep[Origins-Spectral Interpretation-Resource Identification-Security-Regolith Explorer,][]{2011-AGU-Lauretta},
  (162\,173) Ryugu for JAXA Hayabusa2 \citep{2010-COSPAR-38-Yano}, and
  (175\,706) 1996\,FG3 and (341\,843) 2008\,EV5 for the former ESA M3/M4
  candidate MarcoPolo-R \citep{2012-ExA-33-Barucci} and
  \add{ARM \citep[Asteroid Redirect Mission,][]{2015-DPS-Abell},}
  and why the recent proposition for a demonstration project of an asteroid
  deflection by ESA, AIDA \citep[Asteroid Impact \& Deflection Assessment,][]{2012-ESA-AIDA},
  targets the NEA (65\,803) Didymos. 
  This latter point, the protection from asteroid hazard, is certainly
  the most famous aspect of the asteroid research known to the general public,
  and has triggered many initiatives leading to breakthroughs in NEA discovery
  and characterization of their surface and physical properties
  \citep[see, \eg,][among others]{2000-PSS-48-Binzel,
    2000-Icarus-148-Stokes, 
    2002-AsteroidsIII-2.2-Ostro, 
    2004-Icarus-170-Binzel, 
    2007-IAUS-236-Jedicke, 
    2011-ApJ-743-Mainzer,
    2011-AJ-141-Mueller}.\\
  \indent In both attempting to link NEAs and MCs transient
  populations with their source regions and
  meteorites and designing a protection strategy, the study of their
  composition is key.
  Indeed, dynamical studies allows to determine relative
  probabilities of the origin of asteroids belonging to those populations
  \citep[e.g.,][]{2002-Icarus-156-Bottke,
    2012-Icarus-217-Greenstreet}. These links are however not
  sufficient, and must be ascertained by \add{compositional} similarities
  \citep{2008-Nature-454-Vernazza, 2015-AsteroidsIV-Binzel, 2015-AsteroidsIV-Reddy}.
  Moreover, different compositions yield different densities and internal
  structure/cohesion 
  \citep{2012-PSS-73-Carry}, and an asteroid on a impact trajectory with Earth
  of a given size will require a different energy to be deflected or destroyed
  according to its nature
  \citep{jutzi2014hypervelocity}. \\
  \indent Here, we aim at classifying a large number of near-Earth and
  Mars-crosser asteroids into broad
  compositional groups by using imaging archival data.
  We present in Section~\ref{sec: phot} the procedure we used to retrieve the
  photometry at visible wavelengths from the 
  publicly available images of the 
  Sloan Digital Sky Survey (SDSS).
  We describe in Section~\ref{sec: taxo} how we use the SDSS photometry
  to classify the objects into the commonly-used Bus-DeMeo taxonomy of asteroids
  \citep{2009-Icarus-202-DeMeo}, following the work by
  \citet{2013-Icarus-226-DeMeo}.
  We present the results of the classification in
  Section~\ref{sec: result} before discussing their implications for
  source regions in Section~\ref{sec: src} and for surface
  rejuvenation processes in Section~\ref{sec: weather}.

\section{Visible photometry for the Sloan Digital Sky Survey\label{sec: phot}}

  \subsection{The Sloan Digital Sky Survey\label{sec: sdss}}

    \indent The Sloan Digital Sky Survey (SDSS)
    is a wide-field imaging survey
    dedicated to observing galaxies and quasars at different
    wavelengths.
    From 1998 to 2009,
    the survey covered over 14,500 square degrees  
    in 5 filters: \u, \g, \r, \i, \z~
    (centered on 355.1, 468.6, 616.5, 748.1 and 893.1 nm), with estimated
    limiting magnitude of 22.0, 22.2, 22.2, 21.3, and 20.5
    for 95\% completeness \citep{2001-AJ-122-Ivezic}.

  \subsection{The Moving Object Catalog\label{sec: moc}}
    \indent In the course of the survey, 
    \numb{471,569} moving objects were identified in the images and listed in
    the Moving Object Catalogue (SDSS MOC, currently in its 4th
    release, including observations through March 2007).
    Among these, \numb{220,101} were successfully linked to
    \numb{104,449} unique objects 
    corresponding to known asteroids \citep{2001-AJ-122-Ivezic}.
    The remaining \numb{251,468} moving objects listed in the MOC corresponded to
    unknown asteroids at the time of the release (August 2008). \\
    \indent First, we keep objects assigned a number or a provisional
    designation only, \ie, those for which we can retrieve the orbital elements.
    Among these, we select the near-Earth and Mars-crossers asteroids
    according to the limits on their semi-major axis, perihelion, and aphelion
    listed in Table~\ref{tab: orbit}, resulting in
    \numb{2071} observations of \numb{1315} unique objects.
    We then remove observations that are deemed unreliable: with any
    apparent magnitudes greater than the limiting magnitudes reported
    above (Section~\ref{sec: sdss}),
    or any photometric uncertainty greater than 0.05.
    These constraints remove a large portion of the dataset 
    (about \numb{75}\%), primarily due to the larger typical
    error for the \z~filter.
    While there is only a small subset of the sample remaining, we are assured
    of the quality of the data \citep[see][for additional information
      on the definition of photometric cuts]{2013-Icarus-226-DeMeo}.
    Additionally, for higher errors, the
    ambiguity among taxonomic classes possible for an object
    becomes so large that the classification
    (Section~\ref{sec: taxo}) becomes essentially meaningless. 
    In this selection process, we kept 
    \numb{588} observations of \numb{353} individual asteroids from the SDSS
    MOC4, as listed in Table~\ref{tab: orbit}.

\begin{table*}[!t]
\centering
  \begin{tabular}{cccccc cccc}
    \hline
    \hline
    Class & \multicolumn{2}{c}{a (au)} & \multicolumn{2}{c}{q (au)} &
    Q (au) &
    MOC4 & SVO-MOC & SVO$_{\rm griz}$ & SVO$_{\rm gri}$\\
    & min. & max.   &    min. & max.   &    min.   \\
    \hline
    Atens        &  -- & 1.0 &  --   &Q$_{\textrm{\mars}}$& 0.983 &  -- &  -- &  10 &   1 \\
    Apollos      & 1.0 & --  &  --   &  1.017  &   --  &  14 &  18 &  82 &  70 \\
    Amors        &  -- & --  & 1.017 &  1.3    &   --  &  29 &  73 & 111 &  40 \\
    Mars-crosser &--&--&--  &Q$_{\textrm{\mars}}$&   --  & 310 & 383 & 622 & 567 \\
    \hline
    Total &--&--&--&--&--                              & 353 & 474 & 825 & 678 \\
    \hline
  \end{tabular}
  \caption[Definition of NEA classes]{%
    Definition of the dynamical classes of near-Earth and
    Mars-crosser asteroids used in present study, based on the
    semi-major axis (a), perihelion (q), and aphelion (Q) of their orbit.
    All the objects have a perihelion inward of Mars aphelion 
    (Q$_{\textrm{\mars}}$) at 1.666\,au.
    The number of objects
    listed in the SDSS MOC4 (Section~\ref{sec: moc}), 
    identified in SDSS MOC4 using SkyBoT (SVO-MOC, Section~\ref{sec: moc5}), 
    and by the SVO NEA project (Section~\ref{sec: svo})
    are also listed.
    See Fig.~\ref{fig: aei} for the distribution of these objects in the
    orbital element space.
    \label{tab: orbit}
  }
\end{table*}

%
\begin{figure}[!t]
  \includegraphics[width=.49\textwidth]{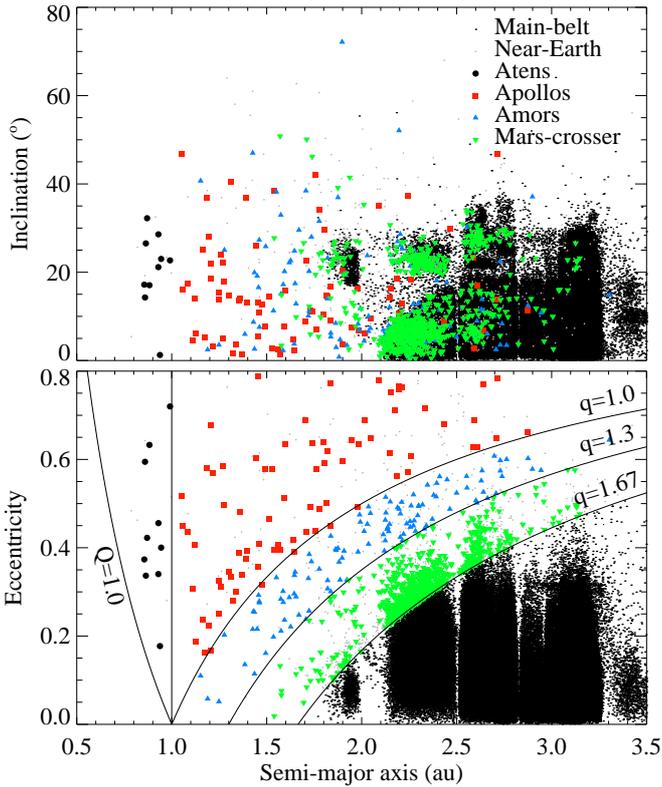}
  \caption[Distribution of NEAs in orbital elements space]{%
    Distribution of the NEAs and Mars-crossers studied here as a function of
    their osculating elements (semi-major axis, eccentricity, and inclination).
    The black dots represent the first 10,000 main-belt asteroids, the grey
    dots the \numb{532} NEAs with spectral classification from the
    literature (see Section~\ref{sec: biblio}), and the
    black circles, red squares, blue triangle, and green stars the
    Atens, Apollos, Amors, and Mars-crossers we classify from SDSS.
    \label{fig: aei}
  }
\end{figure}
%
%

  \subsection{Identifying unknown objects in the MOC4\label{sec: moc5}}
    \indent As mentioned above, more than half of the MOC4 entries
    had not been linked with known asteroids. At the time of the
    release (August 2008), about \numb{460,000} asteroids had been
    discovered and \numb{350,000} were numbered (\ie, had
    well-constrained orbits allowing easy cross-matching with SDSS
    detected sources). 
    The current number of discovered asteroids has now risen above
    700,000, with more than \numb{370,000} numbered objects.
    We therefore use the improved current knowledge on the
    asteroid population to link unknown MOC sources to known
    objects. \\
    \indent We use the Virtual Observatory (VO) SkyBoT cone-search service 
    \citep{2006-ASPC-351-Berthier}, hosted
    at IMCCE\footnote{\href{http://vo.imcce.fr/webservices/skybot/}
      {http://vo.imcce.fr/webservices/skybot/}},
    for that purpose. SkyBoT pre-computes weekly the
    ephemeris of all known Solar System objects for the period
    \numb{1889-2060}, and stores their heliocentric positions with a
    time step of 10 days, allowing fast computation of positions at any
    time. The cone-search tool allows to
    request the list of known objects within a field of view at any
    given epoch as seen from Earth in typically less than 10\,s.
    We send \numb{251,468} requests to SkyBoT, corresponding to the 
    \numb{251,468} unknown objects in the MOC4, centered on the MOC4
    object's coordinates, at the reported epoch of observation, within
    a circular field of view of 30 arcseconds.
    Although many asteroids among the 700,000 known have position
    uncertainty larger than this value (as derived from their orbital
    parameter uncertainty), this cut ensures that we only keep objects
    with a high probability to be linked with the MOC sources.\\
    \indent To further restrict the number of false-positive
    associations, we compare the position, apparent motion, and 
    apparent magnitude of the MOC sources to that predicted from
    ephemeris provided by SkyBoT, based on the database of orbital
    elements
    AstOrb\footnote{\href{http://asteroid.lowell.edu/}{http://asteroid.lowell.edu/}}.
    We consider successful association of SDSS sources with SkyBoT entry if
    the positions are closer than 30\arcsec, the apparent V-Johnson
    magnitudes do not differ by more than 0.5, and the apparent motions are
    co-linear (difference in $d\alpha \cos(\delta)/dt$ and $d\delta/dt$ of
    less than 3\arcsec/h). 
    However, neither SkyBoT nor MOC4 provide estimates on the
    uncertainty in the apparent velocity.
    The only information is the uncertainty in the velocity components
    parallel and perpendicular to the SDSS scanning direction. The mean value
    of this error (both in the parallel and in the perpendicular direction)
    is of 1\arcsec/h. We are taking this value as one standard
    deviation to set the cut above.\\
    Of the \numb{251,468} unidentified MOC sources, SkyBoT
    provides known asteroids within 30 arcseconds for \numb{68,497} 
    (\numb{27}\%), corresponding to \numb{41,055} unique
    asteroids. We trim this value to \numb{57,646} 
    (\numb{36,730} asteroids) for which the association can be considered
    certain.
    The vast majority of these now-identified asteroids have orbits
    within the main belt (\numb{35,404}, corresponding to \numb{96}\%),
    but some are NEAs (\numb{48}, \numb{0.1}\%),
    or MCs (\numb{73}, \numb{0.2}\%).
    Their respective numbers are reported in Table~\ref{tab: orbit}.
    The complete list of MOC entries associated to known asteroids
    (\numb{277,747} entries associated to \numb{141,388} asteroids) is
    freely accessible\footnote{%
      \href{http://svo2.cab.inta-csic.es/vocats/svomoc}
           {http://svo2.cab.inta-csic.es/vocats/svomoc}}.

  \subsection{The SVO Near-Earth Asteroids Recovery Program\label{sec: svo}}
\begin{figure}[!t]
  \includegraphics[width=.49\textwidth]{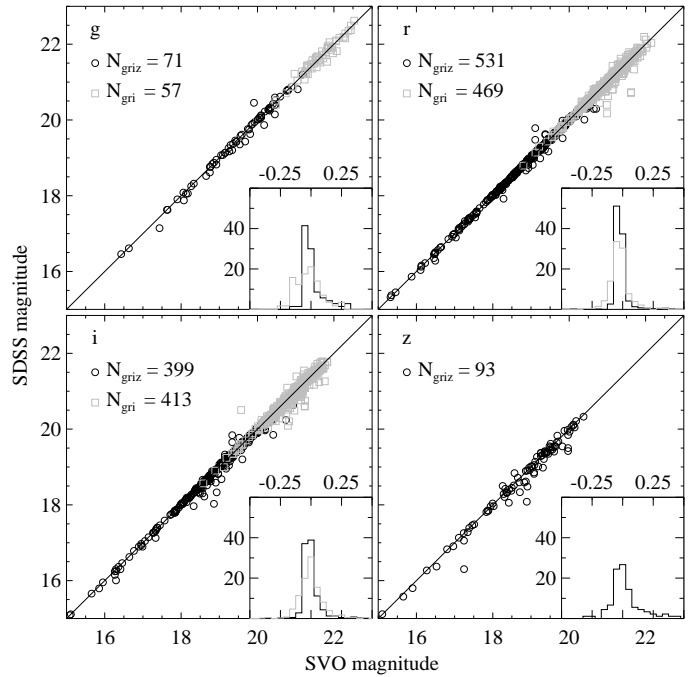}%
  \caption[Comparison of SDSS and SVO apparent magnitude]{%
    Comparison of the apparent magnitude measured by our system (SVO
    magnitude) with that reported in the SDSS MOC4 (SDSS magnitude)
    for the four \g, \r, \i, and \z~filters. For each, the number of
    common objects is reported, for both four-bands (black circles)
    and three-bands (grey squares) sets. 
    The inserted histograms show the distribution of the magnitude
    difference (MOC-SVO)
    \label{fig: appmag}
  }
\end{figure}

    \indent In addition, we search the images of the SDSS for NEAs and
    MCs that were
    either not identified as moving objects by the automatic SDSS pipeline, 
    rejected by the MOC data selection\footnote{%
      \href{http://www.astro.washington.edu/users/ivezic/sdssmoc/sdssmoc.html}
           {http://www.astro.washington.edu/users/ivezic/sdssmoc/sdssmoc.html}},
    or imaged after the latest compilation of the SDSS MOC4 (\ie,
    observed after March 2007).
    Indeed, only moving objects with an apparent motion between
    0.05 and 0.050~deg/day were included in the MOC, leaving a significant
    fraction of NEAs
    un-cataloged \citep{2013-AN--Solano}. \\
    \indent We use the resources of the citizen-science project 
    ``Near-Earth Asteroids Recovery Program''
    of the Spanish Virtual Observatory (SVO) 
    which was originally designed for this very purpose: 
    to identify and measure the astrometry of NEAs in archival imaging data
    \citep{2013-AN--Solano}.
    For each Aten, Amor, Apollo, and Mars-crosser listed by the 
    Minor Planet Center\footnote{%
      \href{http://minorplanetcenter.org/}
           {http://minorplanetcenter.org/}} (MPC), 
    its ephemeris are computed over the period of operation of the
    SDSS imaging survey (1998 to 2009) and compared
    to the footprints of the images of the survey.
    The images possibly containing an object brighter than the SDSS limiting
    magnitude (V\,=\,22) are then proposed to the public for identification
    through a web portal\footnote{%
      \href{http://www.laeff.cab.inta-csic.es/projects/near/main/}
           {http://www.laeff.cab.inta-csic.es/projects/near/main/}}.
    Since the beginning of the project in 2011, over 2,500 astrometry measurements of
    about 600 NEAs not identified in the MOC have been reported to the MPC
    \citep[see][for details on the project]{2013-AN--Solano}. \\
    \indent To compute the photometry of the NEAs measured by the users
    we first searched in the photometric catalog of the 8th SDSS Data
    Release\footnote{\href{http://cdsarc.u-strasbg.fr/viz-bin/Cat?II/306}
      {http://cdsarc.u-strasbg.fr/viz-bin/Cat?II/306}}. If no
    photometry associated with the NEA was found, we ran SExtractor on the
    corresponding images and calibrated the SExtractor magnitudes by
    comparing them with the SDSS magnitudes of the sources identified in the
    image. \\
    \indent Owing to the more stringent limiting magnitude in the
    \z~filter, many asteroids are identified over three bands (\g\r\i) 
    only. We also report these objects here, although deriving a
    taxonomic classification is of course less accurate.
    Overall, we collect
    \numb{1194} \add{four bands} (\g\r\i\z) photometry measurements of \numb{825}
    unique asteroids
    and
    \numb{976} three-bands (\g\r\i) photometry measurements of \numb{678}
    distinct asteroids (Table~\ref{tab: orbit}).
    We present in Fig.~\ref{fig: appmag} a comparison of our
    measurements with the magnitudes reported in the SDSS MOC4 for the
    common asteroids in both sets, showing excellent agreement
    (values agree with a standard deviation of 0.05 mag).

%

\section{Taxonomic classification\label{sec: taxo}}
  \indent The SDSS photometry has been used to classify 
  asteroids according to their colors by many authors
  \citep[\eg,][]{2002-AJ-124-Ivezic, 
    2005-Icarus-173-Nesvorny, 
    2008-Icarus-198-Parker, 
    2010-AA-510-Carvano}. 
  \add{One key advantage of the survey was the almost simultaneous
    acquisition of all filters (5 min in total), hence limiting the
    impact of geometry-related lightcurve on the apparent magnitude.}
  Here we follow the work by \citet{2013-Icarus-226-DeMeo,
    2014-Nature-505-DeMeo}
  in which the class definitions
  are set to be as consistent as possible with previous spectral taxonomies
  based on higher spectral resolution and larger wavelength coverage data sets, 
  specifically Bus and Bus-DeMeo taxonomies \citep{2002-Icarus-158-BusII,
    2009-Icarus-202-DeMeo}.
  We present concisely the classification scheme below and refer to 
  \citet{2013-Icarus-226-DeMeo} for a complete description.

  \subsection{From SDSS to Bus-DeMeo taxonomy\label{sec: sdss2taxo}}
    \indent First, we convert the photometry into reflectance 
     \citep[using solar colors from][]{2006-MNRAS-367-Holmberg} and
    normalize them to unity in filter \g.
    Second, we compute the 
    slope of the continuum over the \g, \r, and \i~filters
    (hereafter gri-slope), and the
    \z$-$\i~color (hereafter zi-color), representing the
    band depth of a potential 1\,$\mu$m band, because they are the most
    characteristic spectral distinguishers in all major taxonomies
    \citep[beginning with][]{1975-Icarus-25-Chapman}. 
    The classification into the taxonomy is then based on these two
    parameters. \\
    \indent As a results of the limited spectral resolution and range of SDSS
    photometry, we group together certain
    classes into broader complexes (see correspondences in 
    Table~\ref{tab: dico}).
    For asteroids with multiple observations that fall under multiple
    classifications, we use the tree-like selection to assign a final
    class 
    (see \citet{2013-Icarus-226-DeMeo} for details and
    \citet{2014-Icarus-229-DeMeo} for an example of a spectroscopic confirmation
    campaign of the SDSS classification used here).
    We successfully classify \numb{982} asteroids from the sample of 
    \numb{1015} near-Earth and Mars-crosser asteroids with four-bands
    photometry (i.e., \numb{97}\% of the sample). \\
    \indent For objects with three-bands photometry
    only, we set their \z~magnitude to the limiting magnitude of 20.5 
    \citep{2001-AJ-122-Ivezic} as an upper limit for their brightness.
    We then classify these asteroids using the scheme presented above.
    Because the magnitude of 20.5 in \z~is an upper limit, the actual zi-color
    for these asteroids may be overestimated. The classification
    can therefore be degenerated, all the classes with similar
    gri-slope and lower zi-color being possible. 
    We assign tentative classification to \numb{254} asteroids from the sample of 
    \numb{678} near-Earth and Mars-crosser asteroids with three-bands
    photometry (i.e., \numb{37}\% of the sample). \\
    \indent In all cases, we mark objects with peculiar spectral behavior with
    the historical notation ``U'' (for unclassified), and discard them from
    the analysis. 
    There are \numb{33} and \numb{424} asteroids in the four-bands and
    three-bands photometry samples respectively for which we \add{cannot}
    assign a class.  
    These figures highlight the ambiguity raised by the lack of
    information on the presence or absence of an absorption band
    around one micron, to which the \z~filter is sensitive.

  \subsection{Gathering classifications from past studies\label{sec: biblio}}
    \indent Many different authors have reported on the taxonomic
    classification of NEAs. We gather here the results of
    \citet{2003-Icarus-163-Dandy},
    \citet{2004-Icarus-170-Binzel},
    \citet{2005-MNRAS-359-Lazzarin},
    \citet{2006-AdSpR-37-deLeon},
    \citet{2010-AA-517-deLeon},
    \citet{2010-Icarus-205-Thomas},
    \citet{2011-AA-535-Popescu},
    \citet{2011-AJ-141-Ye},
    \citet{2011-Icarus-212-Reddy},
    \citet{2012-Icarus-221-Polishook}, 
    \citet{2013-Icarus-225-Sanchez}, and
    \citet{2014-Icarus-227-DeMeo}.
    These authors used different taxonomic schemes to classify their
    observations, using either broad-band filter photometry or
    spectroscopy,  
    at visible wavelengths only or also in the near infrared.
    We therefore transpose the classes of these different schemes
    \citep[][]{1984-PhD-Tholen, 1989-AsteroidsII-Tholen,
      2002-Icarus-158-BusII, 2009-Icarus-202-DeMeo} into the single,
    consistent, set of 10 classes we already use for the SDSS data.
    Here also, we attribute the historical ``U'' designation for
    objects with apparently contradictory classifications
    \citep[\eg, QX or STD in][]{2011-AJ-141-Ye}. 
    These pathological cases represent \numb{15}\% of the objects with
    multiple class determinations.
    In total, we gather \numb{1022} classifications for \numb{648}
    objects listed in the literature.

\begin{table}[t]
  \centering
  \begin{tabular}{ll}
    \hline
    \hline
    SDSS \& Literature & Bus-DeMeo \\
    \hline
{\bf A}, AR, AS                              & A \\
{\bf B}                                      & B \\
{\bf C}, C:, Cb, Cg, Ch                      & C \\
{\bf D}, DT                                  & D \\
{\bf K}, K:                                  & K \\
{\bf L}, Ld                                  & L \\
{\bf Q}, Q/R, R, RQ                          & Q \\
O, Q/R/S, R, RS                              & S \\
{\bf S}, S:, S(I$\rightarrow$V), Sa, Sk, Sl, Sq, Sq:, Sr  & S \\
{\bf V}, V:                                  & V \\
E, M, P, {\bf X}, X:, XT, Xc, Xe, Xk         & X \\
{\bf U}, C(u),  R(u), S(u), ST, STD, QX      & U \\
    \hline
  \end{tabular}
  \caption[Homogeneous taxonomy]{%
    Correspondence between the classes from different taxonomies
    \citep[\eg,][]{1989-AsteroidsII-Tholen, 2002-Icarus-158-BusII}
    found in the literature, used to classify the SDSS photometry (in bold), 
    and their equivalence in the  reduced version of the taxonomy by 
    \citet{2009-Icarus-202-DeMeo} adopted here, following the work by
    \citet{2013-Icarus-226-DeMeo}.
    We strive to preserve the most extreme classes 
    (like A, B) and we convert the tentative classification (\eg, RS)
    into their broader, safer, complex (S).
    We also consider as unknown (U) any data of insufficient quality.
    \label{tab: dico}
  }
\end{table}

%

\section{Results\label{sec: result}}

\begin{figure}[t]
  \includegraphics[width=.49\textwidth]{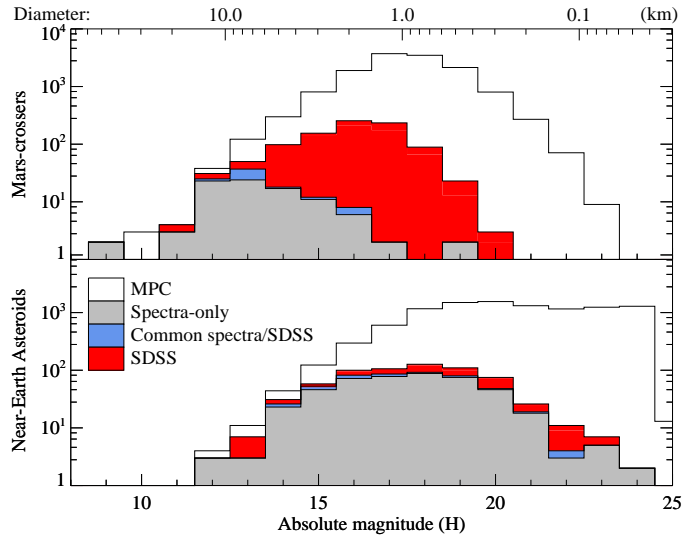}
  \caption[Sample description]{%
    Distribution of the known taxonomy for near-Earth and
    Mars-crosser asteroids as a function of their absolute
    magnitude H. The white histogram corresponds to the total number of
    discoveries (listed in AstOrb), the grey to the taxonomic
    classification found in the literature, 
    the blue to the overlap between literature and present study, and
    the red the four-bands set.
    An approximative conversion to diameter is also reported, 
    using $D = 1329 \times p_v^{-0.5} \times 10^{-0.2H}$, with the geometric
    albedo $p_v$ taken as 0.2 (the majority, $\approx$\numb{60}\%, of the
    objects classified here being S-types). 
    \label{fig: sample}
  }
\end{figure}

  \indent We list in Table~\ref{tab: class4} the photometry and the taxonomy of all
  \numb{982} near-Earth and Mars-crossers asteroids with four-bands photometry. 
  The \numb{254} asteroids with three-bands photometry are listed
  separately in Table~\ref{tab: class3}, because their taxonomic
  classification is less robust.
  In many cases, the upper limit of 20.5 for their \z~magnitude
  provides a weak constraint on their taxonomy, and classes with high
  zi-color (mainly V-types) are more easily identified. This sample
  based on three-bands photometry only is therefore biased, but it
  can be used as a guideline for selecting targets for spectroscopic
  follow-ups.
  We concentrate below on the sample based on four-bands photometry. \\
  \indent The \numb{206} NEAs presented here have absolute magnitudes
  between 12 and 23. Our sample fully overlaps with the size range
  of the \numb{523} NEAs characterized by visible/near-infrared spectroscopy
  published to date and \add{represents} an increase of $\approx$\numb{40}\%
  of the current sample size
  (Fig.~\ref{fig: sample}).
  A significant fraction (46\%) of the NEA population with H\,$<$\,16
  (about 2\,km diameter for an albedo of 0.20)
  has a taxonomic classification. For smaller diameters, the
  fraction drops quickly to 10\% and less. The sub-kilometer
  population of NEAs is therefore still poorly categorized. \\
  \indent The absolute magnitude of the \numb{776} MCs reported here
  ranges from 11 to 19. Our sample represents the first classification of
  sub-kilometric Mars-crossers, and a \numb{sixfold} increase to the sample
  of \numb{117} MCs from spectroscopy (Fig.~\ref{fig: sample}). 
  Similarly to NEAs, about 40\% of the MC population with H\,$<$\,14
  (about 5\,km diameter for an albedo of 0.20)
  \add{now has} a taxonomic classification, and the fraction drops quickly
  to 10\% for smaller diameters.

%
%

    \subsection{Taxonomy and orbital classes}
\begin{figure}[ht]
    \includegraphics[width=.4\textwidth]{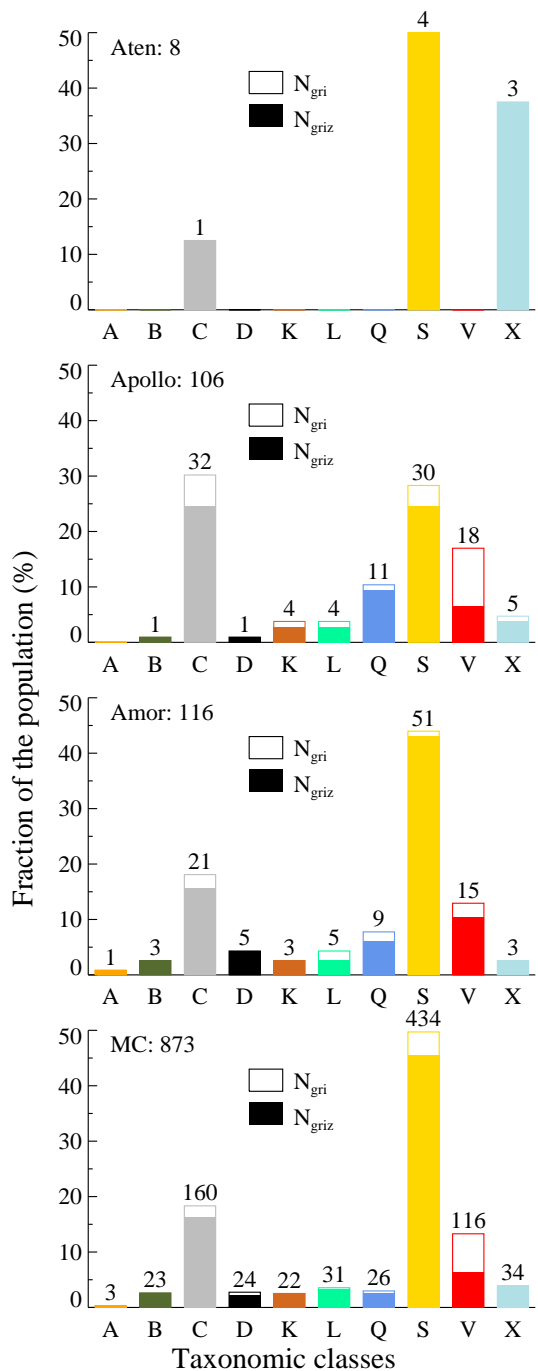}
    \caption[Distribution of classes per orbital group]{%
      Distribution of the taxonomic classes for each
      dynamical group: Aten, Apollo, Amor, and Mars-crosser.
      The number of object with four-band photometry in each class is
      reported.  
      Empty and filled bars stands for the three and four band
      photometry samples respectively.
      \label{fig: taxorb}
    }
  \end{figure}

      \indent We present in Fig.~\ref{fig: taxorb} how the different
      classes distribute among the orbital populations.
      As already reported by \citet{2004-Icarus-170-Binzel}, the
      broad S- (including Q-types), C-, X-complexes, and V-types
      dominate the NEA population, the minor classes (A, B, D, K, L)
      accounting for a few percents only, similarly to what is found
      in the inner main belt \citep{2013-Icarus-226-DeMeo,
        2014-Nature-505-DeMeo}.  
      We find that the S-complex encompasses twice as many objects as
      the C- and X-complexes,
      compared to the threefold difference reported by
      \citet{2004-Icarus-170-Binzel}. \\
%
      \indent The distribution of taxonomic classes among MCs is
      similar to that of Apollos and Amors (with only 8 Atens, our
      sample suffers from low-number statistics).
      Our findings of V-types accounting for roughly \numb{5}\% of all
      MCs may therefore seem puzzling considering 
      the lack of V-types among the $\approx$100 classified MCs
      highlighted by \citet{2004-Icarus-170-Binzel}.
      It is, however, an observation selection effect.
      Indeed, Mars-crossers are inner-main belt
      (IMB) asteroids which eccentricity has been increased by
      numerous weak mean-motion resonances
      \citep[``chaotic diffusion'',
        see][]{1999-Icarus-139-Morbidelli}.
      The IMB hosting the largest reservoir of V-types in the solar
      system \citep[Vestoids,][]{1993-Science-260-Binzel}, V-types 
      were expected in MC space. \\
      \indent The size distribution of known Vestoids however peaks
      at H\,=\,16 (about 1.5\,km diameter), and the largest members
      have an absolute magnitude
      of 14 \citep[we use the list of Vestoids from][]{PDS-Nesvorny}.
      All the V-types identified here have an absolute magnitude above 14,
      and so does (31\,415) 1999 AK23 (H\,=\,14.4), the first V-type
      among MCs reported recently by \citet{2014-PSS-92-Ribeiro}.
      Only 33 MCs had been characterized with this absolute magnitude
      or higher to date, and the previous lack of report of V-types
      among MCs is consistent with our findings. \\
%

    \subsection{Low-$\delta v$ as space mission targets}
      \indent Within the \numb{206} NEA serendipitously observed by the SDSS,
      we identify \numb{36} potential targets for space missions based on
      their accessibility. We select all the NEA with a $\delta v$ below
      \numb{6.5}\,km/s. As a matter of comparison, the required $\delta v$ to
      reach 
      the Moon and Mars are of \numb{6.0} and \numb{6.3}\,km/s
      \citep[e.g.,][]{abell2012near}.
      We list in Table~\ref{tab: deltav} the basic characteristics of these
      potential targets, together with the targets already, or planned to be,
      visited by \add{spacecraft}. \\
      \indent Among the list of low-$\delta v$ objects, we find a large
      majority of S-types, following their dominance in the sample
      presented here of about \numb{60}\%.
      We, however, note the presence of potential D-, L-, and K-types.
      To date, of the 24 taxonomic classes,
      only C- (Mathilde, Ceres),
      S- (Ida, Eros, Gaspra, \add{Itokawa,} and Toutatis),
      Xe- (Steins), 
      Xk- (Lutetia), and
      V-types (Vesta) have been visited by spacecraft.
      These potential D-, L-, and K-types targets may represent good
      opportunities for exploration. 
      Data in the visible can only suggest the presence of an 
      absorption band at 1\,$\mu$m, and near-infrared data is 
      required to confirm these potential classifications.


\begin{table*}[t]
  \centering
  \begin{tabular}{lclccrrccr}
    \hline
    \hline
    Designation & Type & Class & $\delta v$ (km/s) & H &
       \multicolumn{1}{c}{$s$ (\%/100\,nm)} &
       \multicolumn{1}{c}{z-i} &
       \multicolumn{1}{c}{a (au) } &
       \multicolumn{1}{c}{e  } &
       \multicolumn{1}{c}{i (\degr) } \\
    \hline
2004 EU22    & X & Apollo & 4.420 & 23.00 &  0.78 &  0.044 & 1.175 & 0.162 &  5.3 \\
1996 XB27    & D & Amor   & 4.750 & 22.00 &  0.85 &  0.123 & 1.189 & 0.058 &  2.5 \\
2000 TL1     & C & Apollo & 4.870 & 22.00 & -0.09 &  0.002 & 1.338 & 0.300 &  3.6 \\
2001 QC34    & Q & Apollo & 4.970 & 20.00 &  0.69 & -0.228 & 1.128 & 0.187 &  6.2 \\
1999 FN19    & S & Apollo & 5.020 & 22.00 &  0.96 & -0.135 & 1.646 & 0.391 &  2.3 \\
2000 SL10    & Q & Apollo & 5.080 & 21.00 &  0.86 & -0.189 & 1.372 & 0.339 &  1.5 \\
1994 CN2     & S & Apollo & 5.150 & 16.00 &  1.12 & -0.073 & 1.573 & 0.395 &  1.4 \\
2002 LJ3     & S & Amor   & 5.280 & 18.00 &  1.07 & -0.258 & 1.462 & 0.275 &  7.6 \\
2004 UR      & C & Apollo & 5.320 & 22.00 &  0.13 & -0.025 & 1.559 & 0.406 &  2.4 \\
2006 UP      & S & Amor   & 5.350 & 23.00 &  1.43 & -0.074 & 1.586 & 0.301 &  2.3 \\
1994 CC      & S & Apollo & 5.370 & 17.00 &  0.99 & -0.210 & 1.638 & 0.417 &  4.7 \\
2010 WY8     & K & Amor   & 5.670 & 21.00 &  0.92 & -0.042 & 1.385 & 0.136 &  6.0 \\
2002 XP40    & S & Amor   & 5.720 & 19.00 &  1.63 & -0.092 & 1.645 & 0.296 &  3.8 \\
1993 QA      & D & Apollo & 5.740 & 18.00 &  1.02 &  0.174 & 1.476 & 0.315 & 12.6 \\
2004 RK9     & C & Amor   & 5.760 & 21.00 &  0.14 & -0.050 & 1.837 & 0.426 &  6.2 \\
2001 FC7     & C & Amor   & 5.780 & 18.00 &  0.16 & -0.034 & 1.436 & 0.115 &  2.6 \\
1977 VA      & C & Amor   & 5.940 & 19.20 &  0.34 & -0.010 & 1.866 & 0.394 &  3.0 \\
2001 WL15    & C & Amor   & 6.000 & 18.00 &  0.34 & -0.179 & 1.989 & 0.475 &  6.9 \\
2000 XK44    & L & Amor   & 6.080 & 18.00 &  1.24 &  0.020 & 1.724 & 0.385 & 11.2 \\
2003 BH      & V & Apollo & 6.090 & 20.00 &  1.79 & -0.273 & 1.456 & 0.356 & 13.1 \\
2000 NG11    & X & Amor   & 6.130 & 17.00 &  0.27 &  0.089 & 1.881 & 0.368 &  0.8 \\
2000 RW37    & C & Apollo & 6.150 & 20.00 &  0.45 & -0.166 & 1.248 & 0.250 & 13.8 \\
2001 FD90    & V & Amor   & 6.200 & 19.00 &  0.75 & -0.419 & 2.046 & 0.478 &  7.3 \\
2002 PG80    & S & Amor   & 6.210 & 18.00 &  1.09 & -0.225 & 2.013 & 0.438 &  4.4 \\
2004 VB      & S & Apollo & 6.260 & 20.00 &  1.04 & -0.200 & 1.458 & 0.409 & 10.9 \\
1993 DQ1     & S & Amor   & 6.270 & 16.00 &  1.20 & -0.207 & 2.036 & 0.493 & 10.0 \\
2000 YG4     & Q & Amor   & 6.300 & 20.00 &  0.61 & -0.155 & 2.211 & 0.503 &  2.6 \\
2004 KD1     & C & Amor   & 6.330 & 17.00 &  0.13 & -0.110 & 1.720 & 0.331 & 10.1 \\
2004 RS25    & C & Amor   & 6.410 & 20.00 &  0.39 & -0.057 & 2.128 & 0.479 &  6.7 \\
2004 QZ2     & S & Amor   & 6.470 & 18.00 & -5.74 & -0.227 & 2.260 & 0.495 &  1.0 \\
2001 FY      & S & Amor   & 6.530 & 18.00 &  1.36 & -0.132 & 1.886 & 0.327 &  4.7 \\
2009 OC      & S & Amor   & 6.540 & 20.00 &  0.85 & -0.153 & 2.137 & 0.446 &  4.6 \\
2004 XM35    & S & Amor   & 6.560 & 19.00 &  1.13 & -0.111 & 1.837 & 0.301 &  5.4 \\
2005 QG88    & K & Apollo & 6.560 & 20.00 &  0.99 & -0.056 & 1.728 & 0.493 & 11.3 \\
1999 KX4     & V & Amor   & 6.580 & 16.00 &  1.60 & -0.432 & 1.457 & 0.293 & 16.6 \\
2002 TY57    & S & Amor   & 6.600 & 19.00 &  0.86 & -0.154 & 1.922 & 0.327 &  3.5 \\
    \hline
Itokawa      & S & Apollo & 4.632 & 19.20 &&& 1.324 & 0.280 &  1.6 \\ 
Bennu        & C & Apollo & 5.087 & 20.81 &&& 1.126 & 0.204 &  6.0 \\
Ryugu        & B & Apollo & 4.646 & 19.17 &&& 1.189 & 0.190 &  5.9 \\
Didymos      & X & Apollo & 5.098 & 17.94 &&& 1.644 & 0.384 &  3.4 \\
    \hline
  \end{tabular}
  \caption[List of low $\delta v$ targets]{%
    List of NEAs with a low $\delta v$, hence potential targets for
    space missions. For each NEA, the taxonomic class determined here is
    reported, together with the dynamical class, the absolute
    magnitude and the orbital elements.
    The targets of the asteroid-deflection mission AIDA (Didymos)
    and of the return-sample missions
    Hayabusa (Itokawa),
    OSIRIS-REx (Bennu), and
    Hayabusa2 (Ryugu)
    are also included for comparison.
    \label{tab: deltav}
  }
\end{table*}

\section{Source regions\label{sec: src}}
\begin{figure}[!t]
  \includegraphics[width=.49\textwidth]{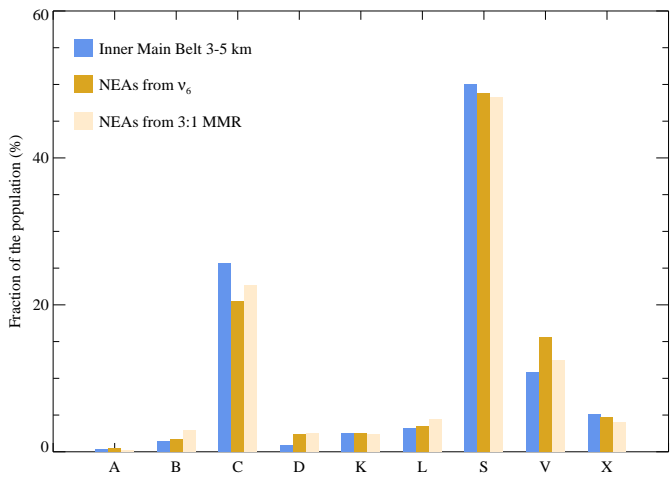}
  \caption[Source region and taxonomic distribution]{%
    Relative fractions (by number) of taxonomic
    classes for asteroids in the inner part of the main-belt (2.0--2.5
    au) with diameter between 3 and 5\,km
    \citep[computed from][]{2013-Icarus-226-DeMeo,
      2014-Nature-505-DeMeo}
    compared with the predicted fractions originating from $\nu_6$ and
    3:1 resonances (see Sec.~\ref{sec: src})
    \label{fig: src}
  }
\end{figure}

  \indent The population of NEAs being eroded on short timescale 
  (\numb{$<$\,10}\,My) by planetary collisions and dynamical ejections,
  new objects must be injected in the NEA space to explain the current
  observed population.
  We use our sample of \numb{982} NEAs and MCs to identify possible source
  regions. For that, we use the source region
  mapper\footnote{Updated by S. Greenstreet from the original mapper
    to include the probability of the source regions of the MCs themselves.} by
  \citet{2012-Icarus-217-Greenstreet}, built on the result of numerical simulations
  of the orbital evolution of test particules in the five regions defined by
  \citet{2002-Icarus-156-Bottke}:
  $\nu_6$ secular resonance, 
  3:1 mean-motion resonance with Jupiter, 
  Mars-crossers (MC), 
  Outer Belt (OB), and
  Jupiter Family Comet (JFC). \\
  \indent For each object, we compute its probability $\mathcal{P}_i$
  to originate from the $i^\textrm{th}$ source region.
  We then normalize all $\mathcal{P}_i$ for each source region.
  The sum of the normalized $\mathcal{P}_i$ over a given taxonomic
  class therefore represents 
  the fraction of objects (by number) of this class in the source region
  (Table~\ref{tab: src}). 
  Uncertainties are computed from the source region mapper
  uncertainties, quadratically added with the margin of 
  error (at 95\% confidence level) to account for the sample size.
  We can then compare these predicted fractions to the observed
  distribution of taxonomic classes for each source region. \\
  \indent The vast majority of NEAs with taxonomic classification sustain diameters of
  less than 5\,km (absolute magnitude above 14). Because the
  distribution of taxonomic classes in the main belt varies with diameter
  \citep{2014-Nature-505-DeMeo}, we need to compare the predicted
  fraction of Table~\ref{tab: src} with objects with H\,$>$\,14 in
  each source region. 
  For the first time, thanks to the large dataset of asteroids
  provided by the Sloan Digital Sky Survey (SDSS), such information is
  available for the inner part of the main belt
  \citep{2013-Icarus-226-DeMeo, 2014-Nature-505-DeMeo} and is reported
  in Table~\ref{tab: src} (3 to 5\,km diameter range). \\
  \indent The comparison of the $\nu_6$ and 3:1 source regions
  (delimiting the inner belt) with the 
  observations in Fig.~\ref{fig: src} shows a very good match of the
  distributions (correlation coefficient of 0.99, maximum difference
  of 5\%). 
  This validates the dynamical path from the inner belt to the
  near-Earth space as described in the model by
  \citet{2012-Icarus-217-Greenstreet}.
  Although the inner belt is widely accepted as a major source for
  NEAs and meteorites
  \citep[e.g.,][]{2002-Icarus-156-Bottke, 2004-Icarus-170-Binzel,
    2008-Nature-454-Vernazza, 2015-AsteroidsIV-Binzel}, 
  the relative contribution of the different source regions particularly with
  respect to asteroid size is still a matter of debate. 
  For the first time, we compare here populations of the same size
  range.

\begin{table*}[ht]
  \centering
  \begin{tabular}{c|r@{\,$\pm$\,}rr@{\,$\pm$\,}rr@{\,$\pm$\,}rr@{\,$\pm$\,}rr@{\,$\pm$\,}r|rr}
    \hline
    \hline
    Class & 
      \multicolumn{10}{c|}{Source regions} &
      \multicolumn{2}{c}{IMB$_{\rm 3-5\,km}$} \\
    &
      \multicolumn{2}{c}{MC} &
      \multicolumn{2}{c}{$\nu_6$} &
      \multicolumn{2}{c}{MMR$_{3:1}$} &
      \multicolumn{2}{c}{OB} &
      \multicolumn{2}{c|}{JFC} &
      (\#) & (\%) \\
    \hline
A   &  0.4 &  5.9 &  0.5 &  6.3 &  0.3 &  5.5 &  0.0 &  3.0 &  0.0 &  3.0 &    9 &  0.4 \\
B   &  2.4 &  5.8 &  1.7 &  5.4 &  2.9 &  6.1 &  6.1 &  7.5 &  0.4 &  4.3 &   34 &  1.5 \\
C   & 15.7 &  5.2 & 20.5 &  5.4 & 22.7 &  5.6 & 38.6 &  6.6 & 47.3 & 10.6 &  589 & 25.7 \\
D   &  2.5 &  5.7 &  2.3 &  5.6 &  2.5 &  5.8 &  5.8 &  7.2 &  9.7 &  8.5 &   20 &  0.9 \\
K   &  2.8 &  5.9 &  2.5 &  5.8 &  2.4 &  5.7 &  4.8 &  6.9 &  1.8 &  7.3 &   57 &  2.5 \\
L   &  3.8 &  5.9 &  3.4 &  5.7 &  4.4 &  6.1 &  1.9 &  5.1 &  0.4 &  4.2 &   73 &  3.2 \\
Q   &  3.6 &  5.6 &  4.3 &  5.8 &  6.3 &  6.4 &  3.8 &  5.8 &  2.4 &  5.5 &    0 &  0.0 \\
S   & 52.8 &  4.6 & 44.5 &  4.6 & 42.0 &  4.8 & 23.8 &  4.8 & 23.4 &  6.9 & 1145 & 50.0 \\
V   & 13.0 &  5.5 & 15.6 &  5.7 & 12.5 &  5.5 &  7.0 &  5.0 &  3.0 &  4.7 &  247 & 10.8 \\
X   &  3.2 &  5.5 &  4.7 &  6.0 &  4.0 &  5.8 &  8.1 &  7.2 & 11.7 & 10.3 &  117 &  5.1 \\
    \hline
  \end{tabular}
  \caption[Source regions]{%
    Relative fraction of each taxonomic class (by number of objects)
    from each source region defined by 
    \citet{2002-Icarus-156-Bottke} and
    \citet{2012-Icarus-217-Greenstreet}, 
    compared with the distribution of taxonomic types among 3 to 5\,km
    inner Main Belt (IMB) asteroids
    \citep[where Q-types are merged with S-types,
      see][]{2013-Icarus-226-DeMeo}. 
    \label{tab: src}
  }
\end{table*}

  \indent Unfortunately, there is no similar data set for the other
  source regions (MC, OB, and JFC) to be compared with our prediction.
  We can still note the overall trend of increasing fraction of C/D/X-types
  in OB and JFC compared with the inner regions, as expected.
  The fraction of K-types peaks in the outer belt, place of the Eos
  family, also as expected, although a steady distribution of K-types
  across source regions is also consistent within uncertainties.
  We also note a strong correlation between MC and small
  inner-belt asteroid populations (correlation coefficient 0.97), which is not
  related to the origin of NEAs, but of MC themselves, via chaotic
  diffusion from the inner belt 
  \citep[see][]{1999-Icarus-139-Morbidelli, 2000-Icarus-145-Michel}. \\
  \indent A peculiar feature of Table~\ref{tab: src} is the high fraction of
  S-type in all source regions. Based on present set of NEAs
  and the dynamical model by \citet{2012-Icarus-217-Greenstreet}, one
  asteroid below 5\,km diameter out of four should be an S-type in the
  outer belt and the same applies to Jupiter family comets. Although the census of
  composition in this size range (3--5\,km) is far from being
  complete in these regions, there is a bias toward detecting S-types
  due to their high albedo ($\approx$0.20) compared to that of
  the C/P/D-types found there 
  \citep[around 0.05, see][for albedo averages over taxonomic classes]{
    2011-ApJ-741-Mainzer, 2013-Icarus-226-DeMeo}
  S-types are minor contributors to the outer belt for diameters above 5\,km
  \citep{2014-Nature-505-DeMeo} and searches for S-type material have
  been unsuccessful among Cybeles, Hildas, and Trojans
  \citep[see][]{2003-Icarus-164-Emery, 2004-Icarus-170-Emery, 2011-AJ-141-Emery, 
        2004-Icarus-172-Fornasier, 2007-Icarus-190-Fornasier, 
        2007-AJ-134-Yang, 2011-AJ-141-Yang, 
        2008-AA-483-Roig, 2008-Icarus-193-Gil-Hutton, 
        2014-AA-568-Marsset}.
  However, only 4\% of the sample (\numb{42} objects) are
  predicted to originate from the OB and JFC source regions. The
  results for these regions is, thus, based on small number statistics.
  The large fraction of S-types in these source regions suggests
  nevertheless that dynamical models may require further refinements.

%

\clearpage
\section{Spectral slope and space weathering\label{sec: weather}}

  \subsection{Size dependence of spectral slope\label{sec:size}}
\begin{figure}[!t]
  \includegraphics[width=.49\textwidth]{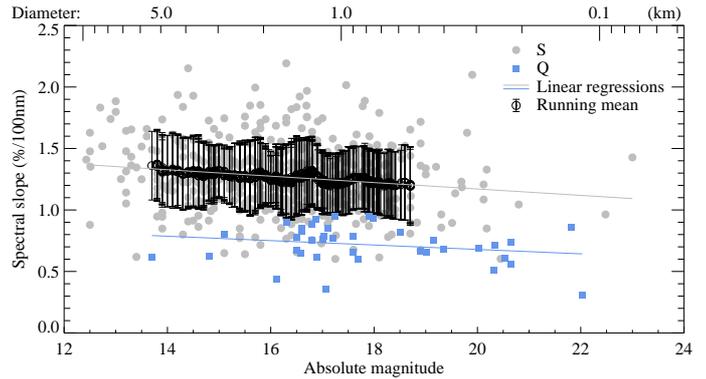}
  \caption[Spectral slope vs diameter]{%
    Spectral slope of 467 S- and 38 Q-types against their absolute magnitude
    (H). An approximative conversion to diameter is also reported (see
    Fig.~\ref{fig: sample}).
    For both, a linear regression is presented, together with a
    running-window average (window size 50) for the sample of S-types.
    \label{fig: HvSL}
  }
\end{figure}

    \indent A size dependency of the spectral characteristics of S-type
    asteroids has been detected early on
    \citep[\eg,][]{1993-Icarus-106-Gaffey} and 
    a correlation between size and spectral slope among NEAs and MCs 
    was shown by \citet{2004-Icarus-170-Binzel} based on a sample of
    \numb{187} S- and \numb{20} Q-types.
    Independently of the other reddening effects due to grain size
    \citep[see, e.g.,][]{2015-Icarus-252-Cloutis} and solar phase angle
    \citep{2002-Icarus-155-Bell, 2012-Icarus-220-Sanchez, 2015-AsteroidsIV-Reddy},
    this trend has been associated with space weathering, 
    causing the spectrum of Q-types, \ie, that of ordinary chondrites
    made of assemblage of olivine and pyroxenes, to \add{redden} into
    S-type spectra under the action of solar wind ions
    \citep[see][among many others]{
      2001-Nature-410-Sasaki, 2005-Icarus-174-Strazzulla}.
    Even if the timescale of this reddening is still debated
    (from $<10^{6}$\,yrs by, e.g., \citet{2009-Nature-458-Vernazza} to
    more than $10^9$\,yrs by, e.g., \citet{2011-Icarus-211-Willman}), it is
    shorter than the age of the Solar System
    \citep[see also][]{2004-Nature-429-Jedicke, 2005-Icarus-173-Nesvorny,
      2008-Icarus-195-Willman, 2010-Icarus-208-Willman, 
      2011-MNRAS-416-DellOro, 2010-ApJ-721-Marchi, 2012-MNRAS-421-Marchi}.
    Some regolith refreshing
    mechanisms must therefore be invoked to explain non- and
    less-weathered surfaces. \\ 
    \indent Because the collisional lifetime decreases with size, 
    the surface is expected to be younger at smaller sizes.
    Smaller asteroids should therefore have a shallower spectral slope
    than larger bodies on average, albeit with a larger dispersion as
    collisions are a stochastic process \citep{2004-Icarus-170-Binzel}.
    This trend has been found recently for the small (diameter
    below 5\,km)
    members of the Koronis family in the main belt
    \citep[$\approx$400 objects,][]{2011-Icarus-211-Rivkin, 2011-Icarus-212-Thomas,
      2012-Icarus-219-Thomas}.
    We present in Fig.~\ref{fig: HvSL} the spectral slope of \numb{467}
    S-types plotted against their absolute magnitude. 
    This sample is almost 3 times bigger than that of
    \citet{2004-Icarus-170-Binzel}, and 
    although we cannot directly compare the values of the spectral
    slope due to the different definitions and normalization
    wavelength,
    we note that we observe the same linear trend from 5 to 1\,km
    \citep[also visible 
      in][]{2012-Icarus-219-Thomas}. However, we do not observe an
    increase of the standard deviation toward smaller diameters nor a
    ``saturation'' regime below 1\,km as visible in their Fig.~7.
    If this trend seems to persist below 1\,km, the statistical
    relevance of this information decreases, however, as the number of objects
    drops dramatically, both here and in \citet{2004-Icarus-170-Binzel}. \\
    \indent We interpret the difference between our findings and those
    reported by \citet{2004-Icarus-170-Binzel} as the effect of the
    larger sample, in which the 
    signal to noise ratio of the data is roughly constant over the
    entire size range: contrarily to the spectral observations,
    generally noisier for small objects, the limiting factor of the
    present SDSS data is the apparent \z~magnitude. Objects with
    four-bands photometry were brighter than magnitude 20 at
    the time of their observations, that is 2 magnitudes brighter than
    the limiting magnitude in \g, \r, and \i, over which the spectral
    slope is computed.\\
    \indent Perhaps not surprisingly, a similar trend is visible for
    the \numb{38} Q-types displayed in Fig.~\ref{fig: HvSL}. These 
    represent the youngest surfaces of their size range.
    Following the argument above, larger asteroids are
    refreshed less often than smaller objects, and this also applies to
    Q-types on their path to redden into S-types, independently of the
    mechanism that originally reset their surface. \\
%

  \subsection{Planetary encounters refresh surfaces\label{sec: encounters}}
%
\begin{figure}[!t]
  \includegraphics[width=.49\textwidth]{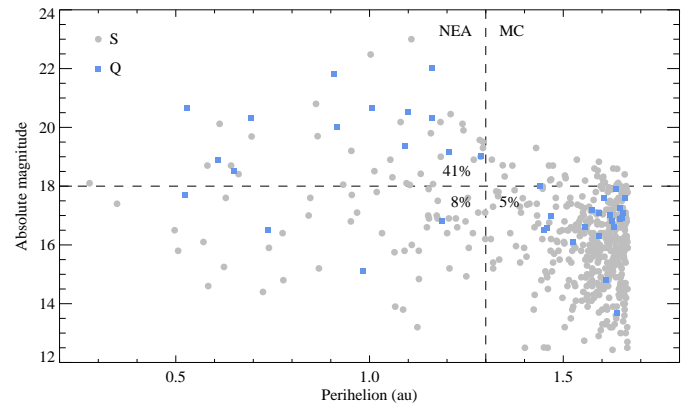}
  \caption[Diameter dependence of Q/S ratio]{%
    Absolute magnitude of S- and Q-types asteroids against perihelion
    distance.
    Below H\,=\,18 (95\% of our MC sample), the Q/S fraction among
    NEAs and MCs is roughly similar (5--8\%).
    The Q/S fraction rapidly increase
    for smaller diameters (H\,$>$\,18) among NEAs, and many
    Q-types will be discovered in the sub-kilometer population of MCs.
    \label{fig: QS}
  }
\end{figure}
%
    \indent If collisions play a stochastic role in modulating the
    spectral slope of silicate-rich asteroids (S-, A-, V-, Q-types), the question
    on the main rejuvenating process is still open.
    \citet{2010-Nature-463-Binzel} and 
    \citet{2014-Icarus-227-DeMeo}  have recently provided
    observational support to the mechanism proposed by
    \citet{2005-Icarus-173-Nesvorny} of 
    close encounters with terrestrial planets.
    The tidal stress during the close encounters has been proposed to
    reveal fresh material (responsible for the Q-type appearance) via
    landslides and regolith shaking. \\
    \indent Both \citet{2010-Nature-463-Binzel} and 
    \citet{2014-Icarus-227-DeMeo} investigated the orbital
    history\footnote{Trajectories of 6 clones per asteroid with
      a difference of 10$^{-6}$ au/yr in initial velocity in each direction were
      integrated with \textsc{swift3 rmvs} \citep{1994-Icarus-108-Levison}
      over 500 kyr, 
      with a time step of 3.65 days. Minimum Orbital Insertion
      Distance (MOID) were averaged over 50 years.}
    of two samples of near-Earth Q- and S-type asteroids,
    searching for planetary encounters
    ``close enough'' (up to a few lunar distances) to reset space
    weathering effect. As a results, all the Q-types
    they tracked had small MOID with either the Earth or
    Mars in the past 500,000 years, a time at which some level of
    space weathering should have 
    already developed (see Sect.~\ref{sec:size} above).
    A significant fraction of S-types had also small MOIDs with
    terrestrial planets. However, the MOID measures the distance
    between two orbits, and not between two bodies, and a small MOID
    does not necessarily implies encounters.
    The authors concluded that planetary encounters, with Mars and the
    Earth, could explain the presence of Q-types among NEAs
    (while they are rare among main-belt asteroids).
    They derived a putative range $r^\star$ of 16 Earth radii at which
    the resurfacing could be felt by asteroids. \\
    \indent Independently, \citet{2010-Icarus-209-Nesvorny} used the sample by
    \citet{2010-Nature-463-Binzel}, using a different approach. By
    tracking\footnote{Trajectories of 100 clones per asteroid,
      spread along the line of variation, were
      integrated also with \textsc{swift3 rmvs} 
      over 1 Myr, 
      with a time step of 1 days.}
    test particles from NEA source regions 
    \citep[similar to][in a way]{2002-Icarus-156-Bottke} to NEA
    space and using a simple step-function model for space weathering,
    they explored the possible range of planetary distances and space
    weathering timescales that would result in the amount 
    and orbital distribution of the Q/S ratio. 
    They concluded on a smaller sphere of influence of planets, with
    $r^\star$ between 5 and 10 Earth radii. Contrarily to
    \citet{2010-Nature-463-Binzel}, who only addressed Earth
    encounters, they found encounters with Venus were as effective as
    those with the Earth. They finally found that encounters with Mars
    were less important, and predicted a very small fraction of
    Q-types among Mars-crossers ($\lesssim$1\%). \\
    \indent The \numb{23} Q-types candidates we identified among
    Mars-crossers
    (Sec.~\ref{sec: result}) account \add{for} about \numb{2.5}\% of the sample 
    (and the Q/S ratio is about \numb{5}\%). 
    This ratio is smaller than for NEAs where it reaches 20\% (the
    total fraction of Q-types among NEAs is 8\%), but it is a lower
    limit.
    The Q/S fraction is indeed strongly diameter-dependent as
    illustrated in Fig.~\ref{fig: QS}. When comparing similar size
    range (H\,$<$\,18, corresponding to 95\% of the MC sample here),
    the Q/S fraction is roughly similar for NEAs
    and MCs, around 5--8\%. The ratio jumps to 40\% for sub-kilometric
    NEAs, and many more Q-types could be discovered among
    sub-kilometric MCs.
    Because of this high fraction of Q-types among MCs, challenging
    the prediction by \citet{2010-Icarus-209-Nesvorny},
    we first derive the theoretical radius of influence during an
    planetary encounter by studying the forces acting on surface grains
    (Sec.~\ref{sec: memo} to Sec.~\ref{sec: distance})
    and then study the dynamical history of
    all the S- and Q-types asteroids presented here, recording their
    close encounters with massive bodies (Sec.~\ref{sec: simu}).

    \subsubsection{Resurfacing model\label{sec: memo}}
      \indent While resurfacing of asteroids by planetary encounters
      \add{has} already been studied
      \citep{2005-Icarus-173-Nesvorny, 2010-Icarus-209-Nesvorny,
        2010-Nature-463-Binzel, 2014-Icarus-227-DeMeo}, the physics of
      the surface was not given much attention in the aforementioned
      articles. The 
      velocity and duration of the encounter, the object's shape,
      internal structure, surface gravity, local slopes, rotation rate
      and orientation, and the nature of the pre-existing regolith and
      its cohesion, were listed as possible parameters dictating the
      distance at which an encounter can resurface the asteroid. \\
      \indent In the following we are
      interested in finding the mean processes responsible for
      resurfacing and the minimum close encounter distances at which
      it would occur.  
      We thus consider a simple force balance equation describing
      the accelerations a surface particle is likely to experience at
      the moment of closest approach. 
      A particle on the asteroid's surface is subject to the following
      forces during a flyby \citep[e.g.,][]{2011-PSS-59-Hartzell} :
      \begin{equation}
        F_{td}+F_{cf}+F_{es}+F_{lt}=F_{ga}+F_{co}+F_{sp} \label{eq:fbal}
      \end{equation}

\begin{figure}[!t]
  \includegraphics[angle=-90, width=.49\textwidth]{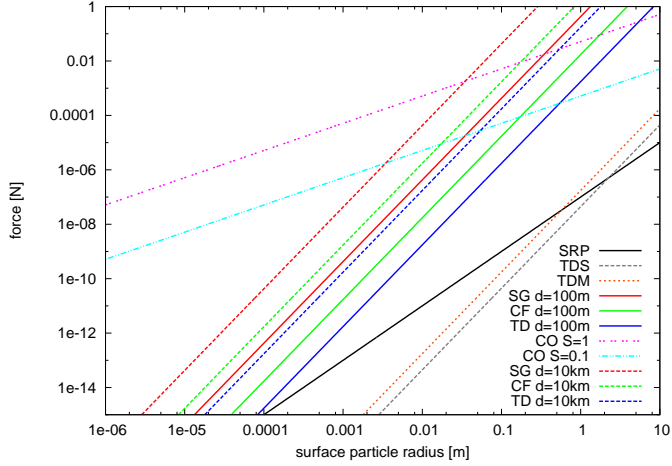}
  \caption[Forces acting on surface particles]{%
    Comparison of the forces acting on surface particles as
    function of their diameter during an encounter with the Earth at
    a 10 Earth-radii distance.
    The different lines represent the absolute magnitude of the
    following forces: self gravity (SG), cohesion (CO), centrifugal
    forces (CF), solar radiation pressure (SRP) for two  
    different asteroid diameters, 100\,m and 10\,km, with the same
    bulk density (1900\,kg.m$^{-3}$) and rotation period ($P$\,=\,10\,h). 
    Tidal accelerations due to the moon (TDM) and the sun (TDS) are
    only visible for the asteroid with a 100\,m diameter. 
    The range of likely cohesive forces is represented by the choice
    of cleanliness parameters between 0.1 and 1
    \citep{perko2001surface, 2010-Icarus-210-Scheeres,
      2011-PSS-59-Hartzell, 2014-MPS-49-Sanchez}.
    One can see that cohesion dominates all other forces for
    particle sizes below $10^{-3}$\,m.
    \label{fig: forces}
  }
\end{figure}

\begin{figure}[!t]
  \includegraphics[width=.49\textwidth]{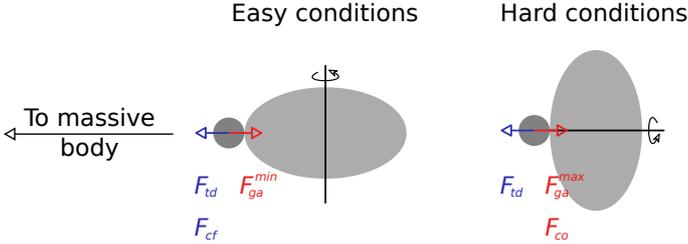}
  \caption[Limit cases for resurfacing event]{%
    Schematic view of the easy
    (\ref{sec:easy}) and hard (\ref{sec:hard}) conditions for
    resurfacing. 
    The forces acting on a surface particles are:
    $F_{td}$ are the tidal forces due to the planetary encounter,
    $F_{cf}$ is the centrifugal force due to the asteroid's rotation,
    $F_{ga}$ is the self-gravity of the asteroid, and
    $F_{co}$ is the cohesion between surface particles.
    \label{fig: easyhard}
  }
\end{figure}

      \indent On the left-hand side of equation (\ref{eq:fbal}) we
      have summed all the forces that can displace the particle:
      $F_{td}$ are the tidal forces due to the planetary encounter,
      $F_{cf}$ is the centrifugal pseudo force due to the asteroid's rotation,
      $F_{es}$ is a repulsive electrostatic force that originates from
      the electric charging of surface particles, and
      $F_{lt}$ is the displacement force acting when the
      asteroid's rotation state changes
      \citep[librational transport, see][]{2014-Icarus-242-Yu}.  
      The right-hand side contains forces that can keep a particle in
      place:
      $F_{ga}$ is the self-gravity of the asteroid,
      $F_{co}$ is the cohesion between surface particles, and
      $F_{sp}$ is the solar radiation pressure.
      We describe the force model in detail in
      \ref{app:forces} and show in Fig.~\ref{fig: forces} their
      absolute magnitude as function of the diameter of surface grains. \\
%
%
%
%
%
      \indent To determine the minimum planet-to-asteroid
      distances that would result in the
      resurfacing of the asteroid, we consider \add{two limiting} cases
      (see Fig.~\ref{fig: easyhard}).
      \textsl{Easy} and \textsl{hard} cases are defined based on
      whether the conditions for resurfacing are favorable or not,
      respectively. By doing so,  we 
      aim at deriving limits to distinguish regions in the parameter space 
      where resurfacing 
      is practically guaranteed, from regions that will leave the
      asteroid's surface untouched. 
      
%
%
    \subsubsection{Easy resurfacing conditions\label{sec:easy}}
      \indent If the asteroid has a rotation rate close to the
      spin-barrier and it passes the planet   
      with its spin axis perpendicular to the orbital plane of the
      hyperbolic encounter, i.e., with zero obliquity with respect to
      the planet, resurfacing is more likely.  
      Since fast rotators are stable with regard to perturbations of
      the spin state, we can assume that the initial spin vector
      remains constant   
      and librational transport will not play a role.
      Low self-gravity is also conducive to resurfacing. The effect of
      solar radiation pressure can be neglected, because  
      it is orders of magnitude weaker than the other
      contributions. \\
      \indent Considering the high porosity of the first layers of
      asteroid surfaces \cite{2012-Icarus-221-Vernazza} that 
      could originate from the electrostatic charging of the surface
      particles, we deem cohesion between surface particles to be
      negligible in this case. 
      Since electrostatic inter-particle repulsion is already
      incorporated in the assumption of a highly porous upper layer of
      regolith,  no additional electrostatic forces shall be considered.
      As a consequence of the above assumptions that yield the best
      case scenario in terms of resurfacing an asteroid,
      equation (\ref{eq:fbal}) simplifies to  
      \begin{equation}
        F_{td}+F_{cf}^{max}=F_{ga}^{min} \label{eq:fbaleasy}
      \end{equation}

      Inserting the forces discussed in \ref{app:forces} into
      equation (\ref{eq:fbaleasy}), 
      we find the distance $r^\star$ between the asteroid and the planet
      where all accelerations cancel. In other words, $r^\star$ is the
      largest planet-to-asteroid distance at which a particle is no
      longer bound to the asteroid.   
      If we assume the asteroid to be a tri-axial ellipsoid defined by
      $R_a \ge R_b \ge R_c$, rearranging
      its surface can take place at a planetary distance of 
      \begin{equation}
        r^\star_{easy}=\left(\frac{6 \mathcal{G} M}{4\pi \mathcal{G}\; \beta\; \rho  - 3 \omega_{sb}^2}\right)^{1/3} \label{eq:rpeasy-tri} \\
      \end{equation}
      \noindent where $\beta=R_b R_c / R_a^2$ is a shape factor relating the
      three principle axes. 
      Equation (\ref{eq:rpeasy-tri}) describes a particle that is on
      the point of the surface farthest away from the center of
      rotation.  
      It is easy to see that for any sort of ellipsoid shape
      the denominator in equation (\ref{eq:rpeasy-tri}) 
      shrinks. Therefore, ellipsoid shapes have extended resurfacing
      distances compared to spherical shapes (in which $\beta$\,=\,1).
      Also, $r^\star_{easy}$ can become arbitrarily large when 
      $\omega=\omega_{sb}=\sqrt{4\pi G \beta \rho / 3}$
      (with $\rho$ the asteroid bulk density), i.e., when the
      asteroid's spin reaches the spin barrier \citep[e.g.,][]{2006-Icarus-181-Pravec}.  
      Since an asteroid is expected to have shed most of its surface material at
      the spin barrier \citep{holsapple2007spin},
      resurfacing will become impossible. Therefore, 
      we will only consider rotation states slightly below this limit
      ($2\pi/\omega=P\sim$\,2.5\,h).

    \subsubsection{Hard resurfacing conditions\label{sec:hard}}
      \indent Resurfacing becomes most difficult, on the other hand,
      if the asteroid has basically no rotation or an obliquity close
      to 90\degr~during its flyby. Then, there is no centrifugal
      acceleration that facilitates the collapse of rubble pile
      columns or lifts particles off the surface. 
      Resurfacing is also more difficult if the asteroid mass
      is high, enhancing its self gravity.  
      If electic charges of surface aggregates are feeble, the particles may
      settle and interlock in dense configurations that are dominated by
      cohesion rather than by electrostatic repulsion 
      \citep{2010-Icarus-210-Scheeres}. 
      Finally, for non-rotating bodies a change in the asteroid
      rotation is likely to occur due to the dynamical instability of
      a non-rotating configuration during flybys.  
      Hence, librational transport could, in principle, occur.
      Yet, since we are interested in the case where resurfacing is
      most difficult, we will neglect its contribution regardless. 
      The equation describing a scenario when resurfacing is most
      difficult thus writes:
      \begin{equation}
        F_{td}=F_{ga}^{max}+F_{co}  \label{eq:fbalhard}
      \end{equation}

      \indent Following the same approach as for easy resurfacing and using
      equation (\ref{eq:fbalhard}) to derive $r^\star_{hard}$ leads 
      to planetary distances that are far below the asteroid's
      disruption regime.  
      Indeed, cohesive forces are dominating all other contributions for
      particle sizes below 10$^{-3}$\,m, as visible in
      Fig.~\ref{fig: forces}, since 99\% of the particles that cover the
      surface have radii below $10^{-4}$\,m (\ref{app:cohesion}). 
      Consequently equation (\ref{eq:fbalhard}) is not a good
      estimator for the resurfacing limit.  
      Since resurfacing is guaranteed when the asteroid enters the
      deformation or even disruption regime, we can simply use
      asteroid's Roche limit as a proxy for the  
      conservative resurfacing distance.
      If we assume a perfectly spherical asteroid without rotation we
      have the following force equilibrium  
      \begin{equation}
        F_{ga}=F_{td} \label{eq:fhard}
      \end{equation}

      \noindent and consequently
      \begin{equation}
        r^\star_{hard}=R\left(\frac{2M}{m}\right)^{1/3} 
                    \approx \left( \frac{3M}{2 \pi \rho}\right)^{1/3} \label{eq:rphard}, \\
      \end{equation}

      \noindent where once again, $R$ is the asteroid's (maximum) radius, and
      $m$ and $M$ are the asteroid and the planet masses. 
      The simple spherical Roche limit serves as a proxy for
      $r^\star_{hard}$, as it would hold as a lower boundary should
      the asteroid be an ellipsoid.

    \subsubsection{Planet to NEA distance for resurfacing\label{sec: distance}}
\begin{figure*}[!t]
  \includegraphics[angle=-90, width=.9\textwidth]{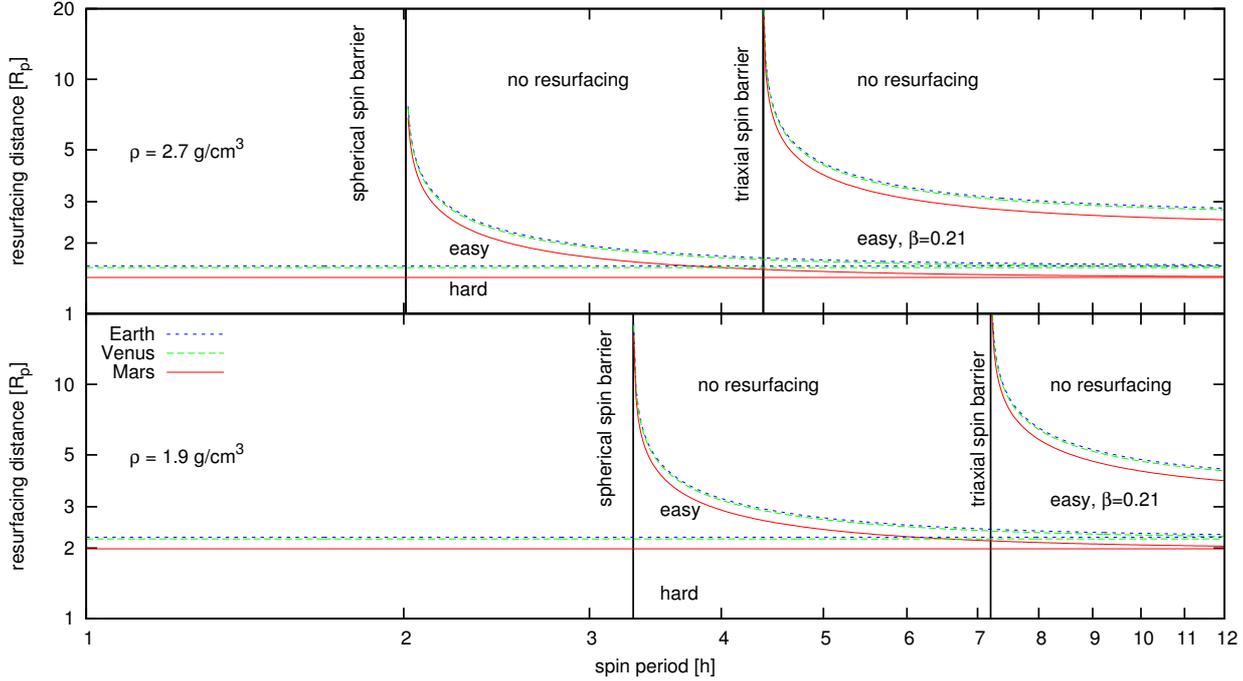}
  \caption[Minimum distance for resurfacing]{%
    Best case ($r^\star_{easy}$) and worst case ($r^\star_{hard}$) resurfacing
    distances in planetary radii for an asteroid (see text for
    details) that has close encounters with Mars,
    Venus and the Earth.
    Resurfacing is possible for close encounter distances below the
    respective lines. Note that forces scale with asteroid
    diameter and this picture is therefore valid for any NEA size.
    \label{fig: spin}
  }
\end{figure*}

      \indent We use the two relations determined above for easy and hard
      resurfacing to evaluate the minimum distance at which a
      planetary encounter can displace surface particles.
      We compute the easy and hard cases \add{for} the following parameters: 
      an absolute magnitude of 20 (corresponding to diameter of 200 and
      600\,m for albedo of 0.5 and 0.05 respectively), 
      a bulk density of 1.9 and 2.7 g.cm$^{-3}$, and a shape factor $\beta$ of
      0.21.
      We show \add{these} threshold distances for Venus, Earth, and Mars,
      as function of the asteroid rotation period in
      Fig.~\ref{fig: spin}. \\
      \indent First, the difference between the easy and hard
      resurfacing distances is substantial, especially for fast
      rotator. Then, we find distances ranging from a couple of
      planetary radii up to 
      10 planetary radii in extreme cases, when the rotation
      period is close to the spin barrier.
      This is fully compatible, yet slightly lower, than the estimates of 5--10 planetary
      radii from \citet{2010-Icarus-209-Nesvorny} derived independently.
      This implies that resurfacing through planetary encounters is
      not common, as the encounters have to be very close. \\
      \indent However, one might ask whether it is permissible to
      simply ignore the effect of cohesion,  
      as has been done in equation \ref{eq:fhard}. In fact, Figure
      \ref{fig: forces} shows that cohesion dominates other forces up
      to millimeter sized particles. 
      However, for larger particles this is no longer the case. If the
      asteroid comes close enough during its approach,  
      tidal forces will then be strong enough to lift decimeter- to
      meter-sized objects on the surface of the asteroid.  
      While objects of this size do not contribute significantly to
      the overall surface of the asteroid \citep{2014-MPS-49-Sanchez},
      they can refresh the surface if displaced by triggering
      landslides for instance. \\
      \indent A similar line of thought
      can be used to argue that we may be able to ignore 
      cohesive and tensile effects in the case of non-spherical
      objects, since the internal stresses decay rapidly towards the
      surface \citep{2007-Icarus-187-Holsapple,2014-MPS-49-Sanchez}. 
      Such an approach would not be permissible, if we were searching
      for criteria to describe  
      global deformation or complete asteroid failure. There, one
      would have to account for internal cohesion and material
      stresses \citep{2006-Icarus-183-Sharma}. 
      Yet, as we are merely interested in whether the combination of
      forces acting   
      during a close encounter can displace any sort of particle on
      the surface of the asteroid, we argue that our simplified
      approach is valid.

%
%
%
%
%
%

    \subsubsection{Dynamical simulations\label{sec: simu}}
\begin{figure*}[t]
    \centering
    \includegraphics[width=.9\textwidth]{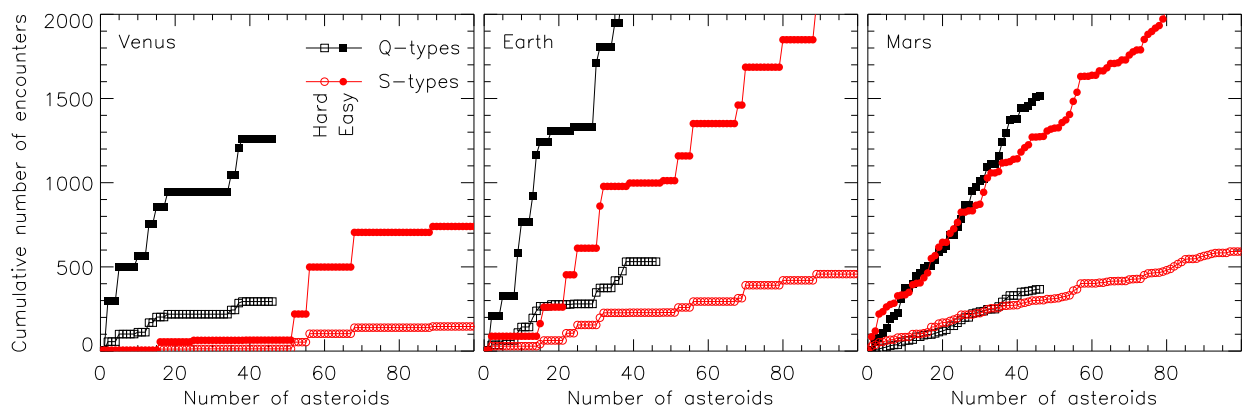}
    \caption[Number of close encounters for S- and Q-types]{%
      Comparison of the number of close encounters for the easy and
      hard resurfacing cases (filled and open symbols) for the samples
      of Q- and S-type asteroids (in black squares and red circles).
      \label{fig: histMOID}
    }
  \end{figure*}

      \indent We investigate the hypothesis of resurfacing by
      planetary encounters developed above by testing whether there is
      a significant difference in the number of potential resurfacing
      events between Q- and S-type asteroid samples. 
      We probe the dynamical history of each asteroid by propagating
      its position together with a sample of 96 clones 0.5 Myrs into the
      past using a symplectic integrator based on Yoshida's T+V split
      \citep{yoshida1990construction}
      with General Relativity (GR) correction
      \citep{lubich2010symplectic}. 
      Symplectic integrators have the advantage that the error in the
      mean anomaly, i.e., the position on the planet on its orbit,
      does not grow as quickly as with standard propagators
      \citep[][and references therein]{eggl2010introduction}.
      Using an 8$^{th}$ order integrator with a stepsize of 1 day in 
      the drift in mean anomaly over 0.5 Myrs is less than
      0.015\degr~in the two-body problem Sun-Earth, corresponding to a
      total along track displacement of less than 6 Earth radii. As
      for close encounters, the propagator is able to resolve all
      close encounters lasting more than 6 hours reasonably well. The
      limit resolution is reached for encounters lasting three
      hours. While the integration algorithm is not regularized, care
      has been taken to avoid losing accuracy due to imprecise 
      calculation of accelerations and the accumulation of round off
      errors (e.g., via the use of Kahan summation). \\
      \indent 
      Each simulation contained the following massive perturbers: the
      8 planets, the
      barycenter of the Pluto system as well the major 
      asteroids (1) Ceres, (2) Pallas, (4) Vesta and (10) Hygiea,
      henceforth referred to as the massive bodies.
      Initial conditions for the massive bodies were taken from JPL DE405
      ephemerides at the epoch J2000. The initial conditions for the asteroids were
      constructed using the open source software
      \texttt{OrbFit}\,\footnote{
        \href{http://adams.dm.unipi.it/orbfit/}{http://adams.dm.unipi.it/orbfit/}}
      \citep{milani2008topocentric}, taking all observations up to April 2014 into
      account.
      Asteroid orbits were fit to the non-relativistic dynamical
      system containing all planets and the Pluto system. No asteroid
      perturbers were taken into account during the fitting
      process. Given the timescales involved in the orbit fitting and
      differential correction, the discrepancies arising 
      from neglecting GR and the major perturbing asteroids on the investigated
      asteroids can be considered small compared to the orbit
      uncertainties. \\
      \indent The
      uncertainty covariance matrix resulting from the orbital fit was then sampled
      along the line of variation \citep{2000-PSS-48-Milani} from
      -3$\sigma$ to +3$\sigma$ with 96 clones 
      per asteroid. The 96 clones were then propagated together with the nominal
      orbit in order to see the dispersion in phase space. All close encounters, with
      the Earth, Venus, Mars, Jupiter and the main perturbing
      asteroids were cataloged for each clone in form of close
      minimum encounter distance (MED) and time histograms. 
      The following limit distances were used to trigger a
      close encounter log:
      Jupiter: 2.56956\,au,
      Venus, Earth and Mars: 0.256956\,au,
      the main perturbing asteroids: 0.0256956\,au.
      These values correspond to approximately 7 Hill's radii
      for Jupiter, 25 for the Earth and Venus, 35 Hill's radii for
      Mars and 35-70 Hill's radii for the perturbing asteroids.
      Minimum encounter distances and velocities
      were calculated using cubic spline interpolation of the asteroid's orbit
      during 
      its close encounter. Global minimum and average encounter distances together
      with their variances were also saved. In addition, minimum orbit intersection
      distances (MOIDs) were calculated every 800 days.
      Similarly to the MED values,
      MOID histograms, averages and variances were cataloged. \\
      \indent 
      It should be mentioned that results stemming from integrating NEO orbits
      backwards in time need to be interpreted with care. Similarly to forward
      propagation NEAs on chaotic trajectories with the terrestrial planets
      have a relatively short horizon that limits an accurate prediction of
      their dynamical evolution
      \citep[see, e.g.,][]{1997-Icarus-129-Michel}
      In such cases even robust ensemble statistics will not yield
      reliable estimates on close encounter distances since the divergence of
      nearby solution becomes exponential. 
      As a consequence, we decided to exclude those NEAs from the statistics
      as soon as the spread in clone encounter distances becomes large.
      In our dynamical study, we also excluded the Yarkovsky
      effect for the following reasons. First, the spin state and orientation
      of the chosen targets are largely unknown. While this could be remedied
      using a statistical distribution of drift parameters among the clones of
      each NEA as proposed, for instance, by 
      \citet{2014-AA-572-Spoto}, this would
      only lead to an artificially increased spread in 96 clones making it
      harder to determine which NEAs are on chaotic orbits and which are
      not. Second, close encounters can change the spin state, 
      making self-consistent predictions very difficult. 
      That being said, Yarkovsky drift rates of our targets range between
      $10^{-4}$ and $10^{-3}$  au/Myr. The cross-track error that results from
      neglecting the Yarkovsky drift alone can range between 2--20 Earth radii
      over 0.5 Myrs, and along track position  errors are much larger. \\
      \indent Regardless of the simplifications of our dynamical
      model, we find that all the Q-types have a minimum MOID allowing
      a close encounter with the Earth (18\% of the sample), or Mars
      (100\%) in the past 500,000 years (we use the median values from
      all the clones here).
      A large fraction of the S-type sample
      also present a minimum MOID that could have led to a close
      encounter with one of the terrestrial planets (12.9\%, 6.7\%, and
      95.4\% for the Earth, Venus, and Mars respectively),
      similarly to the
      situation presented by \citep{2010-Nature-463-Binzel} and
      \citep{2014-Icarus-227-DeMeo}.
      However, a small MOID does not necessary imply a close
      encounter: the MOID provides a measure of the distance between
      the orbits, not between the bodies. The typical example would be
      a Trojan asteroid, for which the MOID is small by definition, but
      that would never encounter the planet.\\
      \indent To overcome this issue, we also study the minimum
      encounter distance (MED) of the asteroids and their clones: that
      is the real distance between the particles and the massive
      bodies. However, even with a symplectic integrator, the drift in
      mean anomaly steadily increases with ephemeris time, and reach
      the level of a few planetary radii, at which the resurfacing is
      deemed to occur (see~\ref{sec: distance} above). Results for MED
      therefore potentially suffer from an underestimation of the
      number of close encounters, especially for the closest.
      The total number of detected MED resurfacing \add{events} is much less
      than the theoretically favorable number of configurations given
      by the encounter numbers using MOIDs. 
      However, MEDs share the same trends with our MOID results.\\
      \indent The analysis using the median properties of each
      asteroid with its 96 clones seems therefore not conclusive: based
      on MED values, the 
      planetary encounters are not expected to play any role in
      rejuvenating the surface of the Q-types, and based on MOID
      values, there should be more fresh surfaces (i.e., less S-types)
      as the vast majority of our sample present small MOID with Mars.
      We therefore study the populations of S- and Q-types
      statistically. We present in Fig.~\ref{fig: histMOID} the 
      cumulative number of close encounters reaching the planetary
      distance for easy and hard resurfacing scenarios (see~\ref{sec:easy}
      and~\ref{sec:hard}) 
      between our sample of asteroids and Venus, the Earth, and Mars.
      There is a clear difference in the number of encounters
      experienced by the population of Q-types compared to the S-types
      sample for both Venus and the Earth. That is, the asteroids
      presenting a fresh surface, similar to ordinary chondrites
      and exempt of signature from space weathering, tend to have more
      close encounters with massive bodies that those asteroids with
      space-weathered surface.
      Significant resurfacing would thus occur from the stress
      during repetitive close encounters with planets.
      The distribution of encounters for both the Earth and Venus
      clearly highlights the sample of Q-types among the NEAs sample
      presented here (and somewhat provides an \textsl{a posteriori}
      validation of our classification into S- and Q-types, the two
      groups being apparently dynamically different).\\
      \indent Conversely, the encounter distribution with Mars is
      similar for both taxonomic classes, as both could theoretically
      have numerous encounters with Mars, within the expected distance
      suitable for resurfacing to occur.
      Among all the parameters recorded during the dynamical
      simulation, none showed a significant difference between the two
      populations and we are therefore limited to speculations.
      Because the sample of MCs studied here exhibit larger
      diameters than the NEAs encountering the Earth, they are deemed to be
      older. Their surface may therefore have reached the 
      saturation state introduced by \citet{2012-MNRAS-421-Marchi},
      where rejunevation is inefficient. The frequent encounters with
      Mars would then no longer turn the S-type surfaces into Q-types.
      Another possibility is that the limit for easy resurfacing being
      tight to the rotation period and obliquity during the close
      encounter, only a small fraction of close encounters do trigger
      resurfacing events, with a preference for fast-spinning
      asteroids. A survey of rotation period of Q-types among NEAs and
      MCs population could address this point.

  \section{Conclusion}
    \indent In this work, we report on the dynamical and surface
    properties of near-Earth asteroids (NEAs) and
    Mars-crosser asteroids (MCs), based on the analysis of their
    colors in the visible. Our sample includes: 

    \begin{itemize}
      \setlength\itemsep{-1ex}
      \item[$\circ$] 43 NEAs and 310 MCs listed in the
        Moving Object Catalogue (MOC4) of the Sloan Digital Sky Survey (SDSS);
      \item[$\circ$]  206
        NEAs and 776 MCs, in publicly available
        images of the SDSS, using our citizen-science project
        ``Near-Earth Asteroids Recovery Program'' of the Spanish
        Virtual Observatory
        \citep[SVO,][]{2013-AN--Solano} measured in four
        filters (\g, \r, \i, and \z);
      \item[$\circ$]   678 NEAs and MCs asteroids measured in three filters (\g,
        \r, and \i) also from our citizen-science project, for 254 of which we assign
        tentative taxonomic classification. 
    \end{itemize}

      In total we have determined the taxonomic class of these 982 NEAs
        and MCs using the DeMeo-Carry taxonomy for SDSS colors
        \citep{2013-Icarus-226-DeMeo}, that is compatible
        with the Bus-DeMeo taxonomy based on V+NIR spectra
        \citep{2009-Icarus-202-DeMeo}

    The sample of taxonomic classes presented here
    correspond to an increase of known classes by 40\% and 600\%
    for NEA and MC populations, respectively.
    Among those, 36 NEAs can be considered 
    potential targets for space missions, owing to their low
    $\delta v$. Some candidates for rare taxonomic classes such as
    D-, L-, and K-types are present within this sample and would benefit from
    further spectral investigations.\\
    \indent We then use the sample of asteroids with taxonomic classes
    based on four-filter observations to study their source regions
    and the effect of planetary encounters on their surface
    properties. To this end

    \begin{itemize}
      \setlength\itemsep{-1ex}
      \item[$\circ$] we compare the distribution of taxonomic
        classes between our sample of NEAs and MCs and the source
        regions, using the predictions resulting from the dynamical
        model presented by \citet{2012-Icarus-217-Greenstreet}. The
        population of 2--5 km diameter asteroids in the main belt
        \citep[][]{2014-Nature-505-DeMeo}
        match closely the predictions from the sample presented here,
        supporting that the $\nu_6$ secular resonance and the 3:1
        mean-motion resonance with Jupiter are the primary sources of
        kilometer-size NEAs;
      \item[$\circ$] we analyze the dependence of spectral slope on
        diameter for asteroids in the S-complex. A linear trend of
        shallower slope toward higher absolute magnitude is found;
      \item[$\circ$] we develop a simple force model on surface grains
        during a planetary encounter; 
      \item[$\circ$] we investigate the planetary 
        distance at which a resurfacing event is expected to occur, by
        considering two extreme cases to the simple force model, when
        the conditions are the most or conversely the least conducive
        to favorable resurfacing event.
        This distance is found to be a function of
        the rotation period and density of the asteroid, and
        of the spin obliquity during the encounters. It ranges from 2
        to $\approx$10 
        planetary radii for Venus, the Earth, and Mars. Such values
        are consistent with previous estimates by
        \citet{2010-Nature-463-Binzel} and
        \citet{2010-Icarus-209-Nesvorny}; 
      \item[$\circ$] we study the dynamical history of the sample of S-
        and Q-type by propagating their positions backward in time for
        0.5 Myrs, together with 96 clones, using a symplectic
        post-Newtonian integrator.
        The population of Q-type presents statistically more
        encounters with Venus and the Earth at distance where
        resurfacing should occur than S-types. However, both
        populations present a high number of encounters with
        Mars and are indistinguishable.
    \end{itemize}

\section*{Acknowledgments}

  \indent We thanks S. Greenstreet for providing the source region
  mapper extended to include Mars-crosser space.
  We acknowledge support from the Faculty of the European
  Space Astronomy Centre (ESAC) for B. Carry's visit.
  S. Eggl would like to acknowledge the support of the European
  Commission H2020-PROTEC-2014 grant no. 640351 (NEOShield-2).
  This publication makes use of the NEAs Precovery Service, developed
  under the Spanish Virtual Observatory project (Centro de
  Astrobiologia, INTA-CSIC) supported from the Spanish MICINN through
  grants AyA2008-02156 and AyA2011-24052. 
  This work was granted access to the HPC resources of MesoPSL financed
  by the Region Ile de France and the project Equip@Meso (reference
  ANR-10-EQPX-29-01) of the programme Investissements d’Avenir supervised
  by the Agence Nationale pour la Recherche.
  This material is based upon work supported, in part, by the National
  Science Foundation under Grant 0907766. Any opinions, findings, and
  conclusions or recommendations expressed in this material are those
  of the authors and do not necessarily reflect the views of the
  National Science Foundation. \\
  \indent This publication makes use of data products from the
  Wide-field Infrared Survey Explorer, which is a joint project
  of the University of California, Los Angeles, and the Jet
  Propulsion Laboratory/California Institute of Technology,
  funded by the National Aeronautics and Space
  Administration. Funding for the creation and distribution of
  the SDSS Archive has been provided by the Alfred P. Sloan
  Foundation, the Participating Institutions, the National
  Aeronautics and Space Administration, the National Science
  Foundation, the U.S. Department of Energy, the Japanese
  Monbukagakusho, and the Max Planck Society. The SDSS Web site
  is \href{http://www.sdss.org/}{http://www.sdss.org/}.


\bibliographystyle{elsarticle-harv}
\bibliography{biblio}

\begin{thebibliography}{109}
\expandafter\ifx\csname natexlab\endcsname\relax\def\natexlab#1{#1}\fi
\expandafter\ifx\csname url\endcsname\relax
  \def\url#1{\texttt{#1}}\fi
\expandafter\ifx\csname urlprefix\endcsname\relax\def\urlprefix{URL }\fi

\bibitem[{Abell et~al.(2012)Abell, Barbee, Mink, Adamo, Alberding, Mazanek,
  Johnson, Yeomans, Chodas, Chamberlin, et~al.}]{abell2012near}
Abell, P., Barbee, B., Mink, R., Adamo, D., Alberding, C., Mazanek, D.,
  Johnson, L., Yeomans, D., Chodas, P., Chamberlin, A., et~al., 2012. The
  near-earth object human space flight accessible targets study (nhats) list of
  near-earth asteroids: identifying potential targets for future exploration.
  In: AAS/Division for Planetary Sciences Meeting Abstracts. Vol.~44.

\bibitem[{{Abell} et~al.(2015){Abell}, {Mazanek}, {Reeves}, {Naasz}, and
  {Cichy}}]{2015-DPS-Abell}
{Abell}, P., {Mazanek}, D., {Reeves}, D., {Naasz}, B., {Cichy}, B., Nov. 2015.
  {NASA's Asteroid Redirect Mission (ARM)}. In: AAS/Division for Planetary
  Sciences Meeting Abstracts. Vol.~47 of AAS/Division for Planetary Sciences
  Meeting Abstracts. p. 312.06.

\bibitem[{{Barucci} et~al.(2012){Barucci}, {Cheng}, {Michel}, {Benner},
  {Binzel}, {Bland}, {B{\"o}hnhardt}, {Brucato}, {Campo Bagatin}, {Cerroni},
  {Dotto}, {Fitzsimmons}, {Franchi}, {Green}, {Lara}, {Licandro}, {Marty},
  {Muinonen}, {Nathues}, {Oberst}, {Rivkin}, {Robert}, {Saladino},
  {Trigo-Rodriguez}, {Ulamec}, and {Zolensky}}]{2012-ExA-33-Barucci}
{Barucci}, M.~A., {Cheng}, A.~F., {Michel}, P., {Benner}, L.~A.~M., {Binzel},
  R.~P., {Bland}, P.~A., {B{\"o}hnhardt}, H., {Brucato}, J.~R., {Campo
  Bagatin}, A., {Cerroni}, P., {Dotto}, E., {Fitzsimmons}, A., {Franchi},
  I.~A., {Green}, S.~F., {Lara}, L.-M., {Licandro}, J., {Marty}, B.,
  {Muinonen}, K., {Nathues}, A., {Oberst}, J., {Rivkin}, A.~S., {Robert}, F.,
  {Saladino}, R., {Trigo-Rodriguez}, J.~M., {Ulamec}, S., {Zolensky}, M., Apr
  2012. {MarcoPolo-R near earth asteroid sample return mission}. Experimental
  Astronomy 33, 645--684.

\bibitem[{{Bell} et~al.(2002){Bell}, {Izenberg}, {Lucey}, {Clark}, {Peterson},
  {Gaffey}, {Joseph}, {Carcich}, {Harch}, {Bell}, {Warren}, {Martin},
  {McFadden}, {Wellnitz}, {Murchie}, {Winter}, {Veverka}, {Thomas}, {Robinson},
  {Malin}, and {Cheng}}]{2002-Icarus-155-Bell}
{Bell}, J.~F., {Izenberg}, N.~I., {Lucey}, P.~G., {Clark}, B.~E., {Peterson},
  C., {Gaffey}, M.~J., {Joseph}, J., {Carcich}, B., {Harch}, A., {Bell}, M.~E.,
  {Warren}, J., {Martin}, P.~D., {McFadden}, L.~A., {Wellnitz}, D., {Murchie},
  S., {Winter}, M., {Veverka}, J., {Thomas}, P., {Robinson}, M.~S., {Malin},
  M., {Cheng}, A., Jan. 2002. {Near-IR Reflectance Spectroscopy of 433 Eros
  from the NIS Instrument on the NEAR Mission. I. Low Phase Angle
  Observations}. Icarus 155, 119--144.

\bibitem[{Berthier et~al.(2006)Berthier, Vachier, Thuillot, Fernique,
  Ochsenbein, Genova, Lainey, and Arlot}]{2006-ASPC-351-Berthier}
Berthier, J., Vachier, F., Thuillot, W., Fernique, P., Ochsenbein, F., Genova,
  F., Lainey, V., Arlot, J., jul 2006. {SkyBoT, a new VO service to identify
  Solar System objects}. In: {C.~Gabriel, C.~Arviset, D.~Ponz, \& S.~Enrique}
  (Ed.), Astronomical Data Analysis Software and Systems XV. Vol. 351 of
  Astronomical Society of the Pacific Conference Series. p. 367.

\bibitem[{Binzel et~al.(2015)Binzel, Reddy, , and
  Dunn}]{2015-AsteroidsIV-Binzel}
Binzel, R., Reddy, V., , Dunn, T., 2015. {The Near-Earth Object Population:
  Connections to Comets, Main-Belt Asteroids, and Meteorites}. Asteroids IV.

\bibitem[{{Binzel}(2000)}]{2000-PSS-48-Binzel}
{Binzel}, R.~P., Apr. 2000. {The Torino Impact Hazard Scale}. Planetary and
  Space Science 48, 297--303.

\bibitem[{{Binzel} et~al.(2010){Binzel}, {Morbidelli}, {Merouane}, {DeMeo},
  {Birlan}, {Vernazza}, {Thomas}, {Rivkin}, {Bus}, and
  {Tokunaga}}]{2010-Nature-463-Binzel}
{Binzel}, R.~P., {Morbidelli}, A., {Merouane}, S., {DeMeo}, F.~E., {Birlan},
  M., {Vernazza}, P., {Thomas}, C.~A., {Rivkin}, A.~S., {Bus}, S.~J.,
  {Tokunaga}, A.~T., Jan. 2010. {Earth encounters as the origin of fresh
  surfaces on near-Earth asteroids}. Nature 463, 331--334.

\bibitem[{Binzel et~al.(2004)Binzel, Rivkin, Stuart, Harris, Bus, and
  Burbine}]{2004-Icarus-170-Binzel}
Binzel, R.~P., Rivkin, A.~S., Stuart, J.~S., Harris, A.~W., Bus, S.~J.,
  Burbine, T.~H., Aug. 2004. {Observed spectral properties of near-Earth
  objects: results for population distribution, source regions, and space
  weathering processes}. Icarus 170, 259--294.

\bibitem[{Binzel and Xu(1993)}]{1993-Science-260-Binzel}
Binzel, R.~P., Xu, S., Apr. 1993. {Chips off of Asteroid 4 Vesta: Evidence for
  the parent body of basaltic achondrite meteorites}. Science 260~(5105),
  186--191.

\bibitem[{Bottke et~al.(2002)Bottke, Morbidelli, {Jedicke}, {Petit}, {Levison},
  {Michel}, and {Metcalfe}}]{2002-Icarus-156-Bottke}
Bottke, W.~F., Morbidelli, A., {Jedicke}, R., {Petit}, J.-M., {Levison}, H.~F.,
  {Michel}, P., {Metcalfe}, T.~S., Apr 2002. {Debiased Orbital and Absolute
  Magnitude Distribution of the Near-Earth Objects}. Icarus 156, 399--433.

\bibitem[{Bus and Binzel(2002)}]{2002-Icarus-158-BusII}
Bus, S.~J., Binzel, R.~P., July 2002. {Phase II of the Small Main-Belt Asteroid
  Spectroscopic Survey: A Feature-Based Taxonomy}. Icarus 158, 146--177.

\bibitem[{{Carry}(2012)}]{2012-PSS-73-Carry}
{Carry}, B., Dec. 2012. {Density of asteroids}. Planetary and Space Science 73,
  98--118.

\bibitem[{Carvano et~al.(2010)Carvano, Hasselmann, Lazzaro, and
  Moth{\'e}-Diniz}]{2010-AA-510-Carvano}
Carvano, J.~M., Hasselmann, H., Lazzaro, D., Moth{\'e}-Diniz, T., 2010.
  {SDSS-based taxonomic classification and orbital distribution of main belt
  asteroids}. Astronomy and Astrophysics 510, A43.

\bibitem[{Chapman et~al.(1975)Chapman, Morrison, and
  Zellner}]{1975-Icarus-25-Chapman}
Chapman, C.~R., Morrison, D., Zellner, B.~H., may 1975. {Surface properties of
  asteroids - A synthesis of polarimetry, radiometry, and spectrophotometry}.
  Icarus 25, 104--130.

\bibitem[{{Cloutis} et~al.(2015){Cloutis}, {Sanchez}, {Reddy}, {Gaffey},
  {Binzel}, {Burbine}, {Hardersen}, {Hiroi}, {Lucey}, {Sunshine}, and
  {Tait}}]{2015-Icarus-252-Cloutis}
{Cloutis}, E.~A., {Sanchez}, J.~A., {Reddy}, V., {Gaffey}, M.~J., {Binzel},
  R.~P., {Burbine}, T.~H., {Hardersen}, P.~S., {Hiroi}, T., {Lucey}, P.~G.,
  {Sunshine}, J.~M., {Tait}, K.~T., May 2015. {Olivine-metal mixtures: Spectral
  reflectance properties and application to asteroid reflectance spectra}.
  Icarus 252, 39--82.

\bibitem[{{Dandy} et~al.(2003){Dandy}, {Fitzsimmons}, and
  {Collander-Brown}}]{2003-Icarus-163-Dandy}
{Dandy}, C.~L., {Fitzsimmons}, A., {Collander-Brown}, S.~J., Jun. 2003.
  {Optical colors of 56 near-Earth objects: trends with size and orbit}. Icarus
  163, 363--373.

\bibitem[{{de Le{\'o}n} et~al.(2006){de Le{\'o}n}, {Licandro}, {Duffard}, and
  {Serra-Ricart}}]{2006-AdSpR-37-deLeon}
{de Le{\'o}n}, J., {Licandro}, J., {Duffard}, R., {Serra-Ricart}, M., 2006.
  {Spectral analysis and mineralogical characterization of 11 olivine pyroxene
  rich NEAs}. Advances in Space Research 37, 178--183.

\bibitem[{{de Le{\'o}n} et~al.(2010){de Le{\'o}n}, {Licandro}, {Serra-Ricart},
  {Pinilla-Alonso}, and {Campins}}]{2010-AA-517-deLeon}
{de Le{\'o}n}, J., {Licandro}, J., {Serra-Ricart}, M., {Pinilla-Alonso}, N.,
  {Campins}, H., Jul. 2010. {Observations, compositional, and physical
  characterization of near-Earth and Mars-crosser asteroids from a
  spectroscopic survey}. Astronomy and Astrophysics 517, A23.

\bibitem[{{Dell'Oro} et~al.(2011){Dell'Oro}, {Marchi}, and
  {Paolicchi}}]{2011-MNRAS-416-DellOro}
{Dell'Oro}, A., {Marchi}, S., {Paolicchi}, P., Sep. 2011. {Collisional
  evolution of near-Earth asteroids and refreshing of the space-weathering
  effects}. Monthly Notices of the Royal Astronomical Society 416, L26--L30.

\bibitem[{{DeMeo} and {Carry}(2013)}]{2013-Icarus-226-DeMeo}
{DeMeo}, F., {Carry}, B., Jul 2013. {The taxonomic distribution of asteroids
  from multi-filter all-sky photometric surveys}. Icarus 226, 723--741.

\bibitem[{{DeMeo} et~al.(2014{\natexlab{a}}){DeMeo}, {Binzel}, {Carry},
  {Polishook}, and {Moskovitz}}]{2014-Icarus-229-DeMeo}
{DeMeo}, F.~E., {Binzel}, R.~P., {Carry}, B., {Polishook}, D., {Moskovitz},
  N.~A., Feb. 2014{\natexlab{a}}. {Unexpected D-type interlopers in the inner
  main belt}. Icarus 229, 392--399.

\bibitem[{{DeMeo} et~al.(2014{\natexlab{b}}){DeMeo}, {Binzel}, and
  {Lockhart}}]{2014-Icarus-227-DeMeo}
{DeMeo}, F.~E., {Binzel}, R.~P., {Lockhart}, M., Jan. 2014{\natexlab{b}}. {Mars
  encounters cause fresh surfaces on some near-Earth asteroids}. Icarus 227,
  112--122.

\bibitem[{DeMeo et~al.(2009)DeMeo, Binzel, Slivan, and
  Bus}]{2009-Icarus-202-DeMeo}
DeMeo, F.~E., Binzel, R.~P., Slivan, S.~M., Bus, S.~J., jul 2009. {An extension
  of the Bus asteroid taxonomy into the near-infrared}. Icarus 202, 160--180.

\bibitem[{{DeMeo} and {Carry}(2014)}]{2014-Nature-505-DeMeo}
{DeMeo}, F.~E., {Carry}, B., Jan. 2014. {Solar System evolution from
  compositional mapping of the asteroid belt}. Nature 505, 629--634.

\bibitem[{Eggl and Dvorak(2010)}]{eggl2010introduction}
Eggl, S., Dvorak, R., 2010. An introduction to common numerical integration
  codes used in dynamical astronomy. In: Dynamics of small solar system bodies
  and exoplanets. Springer, pp. 431--480.

\bibitem[{{Emery} and {Brown}(2003)}]{2003-Icarus-164-Emery}
{Emery}, J.~P., {Brown}, R.~H., Jul. 2003. {Constraints on the surface
  composition of Trojan asteroids from near-infrared (0.8-4.0 {$\mu$}m)
  spectroscopy}. Icarus 164, 104--121.

\bibitem[{{Emery} and {Brown}(2004)}]{2004-Icarus-170-Emery}
{Emery}, J.~P., {Brown}, R.~H., Jul. 2004. {The surface composition of Trojan
  asteroids: constraints set by scattering theory}. Icarus 170, 131--152.

\bibitem[{Emery et~al.(2011)Emery, Burr, and Cruikshank}]{2011-AJ-141-Emery}
Emery, J.~P., Burr, D.~M., Cruikshank, D.~P., Jan. 2011. {Near-infrared
  Spectroscopy of Trojan Asteroids: Evidence for Two Compositional Groups}.
  Astronomical Journal 141, 25.

\bibitem[{{Fornasier} et~al.(2007){Fornasier}, {Dotto}, {Hainaut}, {Marzari},
  {Boehnhardt}, {de Luise}, and {Barucci}}]{2007-Icarus-190-Fornasier}
{Fornasier}, S., {Dotto}, E., {Hainaut}, O., {Marzari}, F., {Boehnhardt}, H.,
  {de Luise}, F., {Barucci}, M.~A., Oct. 2007. {Visible spectroscopic and
  photometric survey of Jupiter Trojans: Final results on dynamical families}.
  Icarus 190, 622--642.

\bibitem[{{Fornasier} et~al.(2004){Fornasier}, {Dotto}, {Marzari}, {Barucci},
  {Boehnhardt}, {Hainaut}, and {de Bergh}}]{2004-Icarus-172-Fornasier}
{Fornasier}, S., {Dotto}, E., {Marzari}, F., {Barucci}, M.~A., {Boehnhardt},
  H., {Hainaut}, O., {de Bergh}, C., Nov. 2004. {Visible spectroscopic and
  photometric survey of L5 Trojans: investigation of dynamical families}.
  Icarus 172, 221--232.

\bibitem[{Fujiwara et~al.(2006)Fujiwara, Kawaguchi, Yeomans, Abe, Mukai, Okada,
  Saito, Yano, Yoshikawa, Scheeres, Barnouin-Jha, Cheng, Demura, Gaskell,
  Hirata, Ikeda, Kominato, Miyamoto, Nakamura, Sasaki, and
  Uesugi}]{2006-Science-312-Fujiwara}
Fujiwara, A., Kawaguchi, J., Yeomans, D.~K., Abe, M., Mukai, T., Okada, T.,
  Saito, J., Yano, H., Yoshikawa, M., Scheeres, D.~J., Barnouin-Jha, O.~S.,
  Cheng, A.~F., Demura, H., Gaskell, G.~W., Hirata, N., Ikeda, H., Kominato,
  T., Miyamoto, H., Nakamura, R., Sasaki, S., Uesugi, K., 2006. {The
  Rubble-Pile Asteroid Itokawa as Observed by Hayabusa}. Science 312,
  1330--1334.

\bibitem[{{Gaffey} et~al.(1993){Gaffey}, {Burbine}, {Piatek}, {Reed}, {Chaky},
  {Bell}, and {Brown}}]{1993-Icarus-106-Gaffey}
{Gaffey}, M.~J., {Burbine}, T.~H., {Piatek}, J.~L., {Reed}, K.~L., {Chaky},
  D.~A., {Bell}, J.~F., {Brown}, R.~H., Dec. 1993. {Mineralogical variations
  within the S-type asteroid class}. Icarus 106, 573--602.

\bibitem[{{Gil-Hutton} and {Brunini}(2008)}]{2008-Icarus-193-Gil-Hutton}
{Gil-Hutton}, R., {Brunini}, A., Feb. 2008. {Surface composition of Hilda
  asteroids from the analysis of the Sloan Digital Sky Survey colors}. Icarus
  193, 567--571.

\bibitem[{{Gounelle} et~al.(2006){Gounelle}, {Spurn{\'y}}, and
  {Bland}}]{2006-MPS-41-Gounelle}
{Gounelle}, M., {Spurn{\'y}}, P., {Bland}, P.~A., Jan. 2006. {The orbit and
  atmospheric trajectory of the Orgueil meteorite from historical records}.
  Meteoritics and Planetary Science 41, 135--150.

\bibitem[{{Greenstreet} et~al.(2012){Greenstreet}, {Ngo}, and
  {Gladman}}]{2012-Icarus-217-Greenstreet}
{Greenstreet}, S., {Ngo}, H., {Gladman}, B., Jan. 2012. {The orbital
  distribution of Near-Earth Objects inside Earth's orbit}. Icarus 217,
  355--366.

\bibitem[{{Hartzell} and {Scheeres}(2011)}]{2011-PSS-59-Hartzell}
{Hartzell}, C.~M., {Scheeres}, D.~J., Nov. 2011. {The role of cohesive forces
  in particle launching on the Moon and asteroids}. Planetary and Space Science
  59, 1758--1768.

\bibitem[{{Holmberg} et~al.(2006){Holmberg}, {Flynn}, and
  {Portinari}}]{2006-MNRAS-367-Holmberg}
{Holmberg}, J., {Flynn}, C., {Portinari}, L., Apr. 2006. {The colours of the
  Sun}. MNRAS 367, 449--453.

\bibitem[{Holsapple(2007)}]{holsapple2007spin}
Holsapple, K.~A., 2007. Spin limits of solar system bodies: From the small
  fast-rotators to 2003 el61. Icarus 187~(2), 500--509.

\bibitem[{{Holsapple}(2007)}]{2007-Icarus-187-Holsapple}
{Holsapple}, K.~A., Apr. 2007. {Spin limits of Solar System bodies: From the
  small fast-rotators to 2003 EL61}. Icarus 187, 500--509.

\bibitem[{{Ivezi{\'c}} et~al.(2002){Ivezi{\'c}}, {Lupton}, {Juri{\'c}},
  {Tabachnik}, {Quinn}, {Gunn}, {Knapp}, {Rockosi}, and
  {Brinkmann}}]{2002-AJ-124-Ivezic}
{Ivezi{\'c}}, {\v Z}., {Lupton}, R.~H., {Juri{\'c}}, M., {Tabachnik}, S.,
  {Quinn}, T., {Gunn}, J.~E., {Knapp}, G.~R., {Rockosi}, C.~M., {Brinkmann},
  J., Nov. 2002. {Color Confirmation of Asteroid Families}. Astronomical
  Journal 124, 2943--2948.

\bibitem[{{Ivezi{\'c}} et~al.(2001){Ivezi{\'c}}, {Tabachnik}, {Rafikov},
  {Lupton}, {Quinn}, {Hammergren}, {Eyer}, {Chu}, {Armstrong}, {Fan},
  {Finlator}, {Geballe}, {Gunn}, {Hennessy}, {Knapp}, {Leggett}, {Munn},
  {Pier}, {Rockosi}, {Schneider}, {Strauss}, {Yanny}, {Brinkmann}, {Csabai},
  {Hindsley}, {Kent}, {Lamb}, {Margon}, {McKay}, {Smith}, {Waddel}, {York}, and
  {the SDSS Collaboration}}]{2001-AJ-122-Ivezic}
{Ivezi{\'c}}, {\v Z}., {Tabachnik}, S., {Rafikov}, R., {Lupton}, R.~H.,
  {Quinn}, T., {Hammergren}, M., {Eyer}, L., {Chu}, J., {Armstrong}, J.~C.,
  {Fan}, X., {Finlator}, K., {Geballe}, T.~R., {Gunn}, J.~E., {Hennessy},
  G.~S., {Knapp}, G.~R., {Leggett}, S.~K., {Munn}, J.~A., {Pier}, J.~R.,
  {Rockosi}, C.~M., {Schneider}, D.~P., {Strauss}, M.~A., {Yanny}, B.,
  {Brinkmann}, J., {Csabai}, I., {Hindsley}, R.~B., {Kent}, S., {Lamb}, D.~Q.,
  {Margon}, B., {McKay}, T.~A., {Smith}, J.~A., {Waddel}, P., {York}, D.~G.,
  {the SDSS Collaboration}, Nov. 2001. {Solar System Objects Observed in the
  Sloan Digital Sky Survey Commissioning Data}. Astronomical Journal 122,
  2749--2784.

\bibitem[{Jedicke et~al.(2007)Jedicke, Magnier, Kaiser, and
  Chambers}]{2007-IAUS-236-Jedicke}
Jedicke, R., Magnier, E.~A., Kaiser, N., Chambers, K.~C., 2007. {The next
  decade of Solar System discovery with Pan-STARRS}. In: Valsecchi, G.~B.,
  Vokrouhlick{\'y}, D., Milani, A. (Eds.), IAU Symposium. Vol. 236 of IAU
  Symposium. pp. 341--352.

\bibitem[{{Jedicke} et~al.(2004){Jedicke}, {Nesvorn{\'y}}, {Whiteley},
  {Ivezi{\'c}}, and {Juri{\'c}}}]{2004-Nature-429-Jedicke}
{Jedicke}, R., {Nesvorn{\'y}}, D., {Whiteley}, R., {Ivezi{\'c}}, {\v Z}.,
  {Juri{\'c}}, M., May 2004. {An age-colour relationship for main-belt
  S-complex asteroids}. Nature 429, 275--277.

\bibitem[{Jutzi and Michel(2014)}]{jutzi2014hypervelocity}
Jutzi, M., Michel, P., 2014. Hypervelocity impacts on asteroids and momentum
  transfer i. numerical simulations using porous targets. Icarus 229, 247--253.

\bibitem[{{Lauretta} et~al.(2011){Lauretta}, {Drake}, and
  {Team}}]{2011-AGU-Lauretta}
{Lauretta}, D.~S., {Drake}, M.~J., {Team}, O., Dec. 2011. {OSIRIS-REx -
  Exploration of Asteroid (101955) 1999 RQ36}. AGU Fall Meeting Abstracts.

\bibitem[{{Lazzarin} et~al.(2005){Lazzarin}, {Marchi}, {Magrin}, and
  {Licandro}}]{2005-MNRAS-359-Lazzarin}
{Lazzarin}, M., {Marchi}, S., {Magrin}, S., {Licandro}, J., Jun. 2005.
  {Spectroscopic investigation of near-Earth objects at Telescopio Nazionale
  Galileo}. Monthly Notices of the Royal Astronomical Society 359, 1575--1582.

\bibitem[{{Levison} and {Duncan}(1994)}]{1994-Icarus-108-Levison}
{Levison}, H.~F., {Duncan}, M.~J., Mar. 1994. {The long-term dynamical behavior
  of short-period comets}. Icarus 108, 18--36.

\bibitem[{Lubich et~al.(2010)Lubich, Walther, and
  Br{\"u}gmann}]{lubich2010symplectic}
Lubich, C., Walther, B., Br{\"u}gmann, B., 2010. Symplectic integration of
  post-newtonian equations of motion with spin. Physical Review D 81~(10),
  104025.

\bibitem[{{Mainzer} et~al.(2011{\natexlab{a}}){Mainzer}, {Grav}, {Bauer},
  {Masiero}, {McMillan}, {Cutri}, {Walker}, {Wright}, {Eisenhardt}, {Tholen},
  {Spahr}, {Jedicke}, {Denneau}, {DeBaun}, {Elsbury}, {Gautier}, {Gomillion},
  {Hand}, {Mo}, {Watkins}, {Wilkins}, {Bryngelson}, {Del Pino Molina}, {Desai},
  {G{\'o}mez Camus}, {Hidalgo}, {Konstantopoulos}, {Larsen}, {Maleszewski},
  {Malkan}, {Mauduit}, {Mullan}, {Olszewski}, {Pforr}, {Saro}, {Scotti}, and
  {Wasserman}}]{2011-ApJ-743-Mainzer}
{Mainzer}, A., {Grav}, T., {Bauer}, J., {Masiero}, J., {McMillan}, R.~S.,
  {Cutri}, R.~M., {Walker}, R., {Wright}, E., {Eisenhardt}, P., {Tholen},
  D.~J., {Spahr}, T., {Jedicke}, R., {Denneau}, L., {DeBaun}, E., {Elsbury},
  D., {Gautier}, T., {Gomillion}, S., {Hand}, E., {Mo}, W., {Watkins}, J.,
  {Wilkins}, A., {Bryngelson}, G.~L., {Del Pino Molina}, A., {Desai}, S.,
  {G{\'o}mez Camus}, M., {Hidalgo}, S.~L., {Konstantopoulos}, I., {Larsen},
  J.~A., {Maleszewski}, C., {Malkan}, M.~A., {Mauduit}, J.-C., {Mullan}, B.~L.,
  {Olszewski}, E.~W., {Pforr}, J., {Saro}, A., {Scotti}, J.~V., {Wasserman},
  L.~H., Dec 2011{\natexlab{a}}. {NEOWISE Observations of Near-Earth Objects:
  Preliminary Results}. Astrophysical Journal 743, 156.

\bibitem[{{Mainzer} et~al.(2011{\natexlab{b}}){Mainzer}, {Grav}, {Masiero},
  {Hand}, {Bauer}, {Tholen}, {McMillan}, {Spahr}, {Cutri}, {Wright}, {Watkins},
  {Mo}, and {Maleszewski}}]{2011-ApJ-741-Mainzer}
{Mainzer}, A., {Grav}, T., {Masiero}, J., {Hand}, E., {Bauer}, J., {Tholen},
  D., {McMillan}, R.~S., {Spahr}, T., {Cutri}, R.~M., {Wright}, E., {Watkins},
  J., {Mo}, W., {Maleszewski}, C., Nov. 2011{\natexlab{b}}. {NEOWISE Studies of
  Spectrophotometrically Classified Asteroids: Preliminary Results}.
  Astrophysical Journal 741, 90.

\bibitem[{{Marchi} et~al.(2010){Marchi}, {De Sanctis}, {Lazzarin}, and
  {Magrin}}]{2010-ApJ-721-Marchi}
{Marchi}, S., {De Sanctis}, M.~C., {Lazzarin}, M., {Magrin}, S., Oct. 2010. {On
  the Puzzle of Space Weathering Alteration of Basaltic Asteroids}.
  Astrophysical Journal 721, L172--L176.

\bibitem[{{Marchi} et~al.(2012){Marchi}, {Paolicchi}, and
  {Richardson}}]{2012-MNRAS-421-Marchi}
{Marchi}, S., {Paolicchi}, P., {Richardson}, D.~C., Mar. 2012. {Collisional
  evolution and reddening of asteroid surfaces - I. The problem of conflicting
  time-scales and the role of size-dependent effects}. Monthly No 421, 2--8.

\bibitem[{{Marsset} et~al.(2014){Marsset}, {Vernazza}, {Gourgeot}, {Dumas},
  {Birlan}, {Lamy}, and {Binzel}}]{2014-AA-568-Marsset}
{Marsset}, M., {Vernazza}, P., {Gourgeot}, F., {Dumas}, C., {Birlan}, M.,
  {Lamy}, P., {Binzel}, R.~P., Aug. 2014. {Similar origin for low- and
  high-albedo Jovian Trojans and Hilda asteroids?} Astronomy and Astrophysics
  568, L7.

\bibitem[{{McSween} et~al.(2006){McSween}, Lauretta, and
  Leshin}]{2006-MESS2-McSween}
{McSween}, H.~Y., Lauretta, D.~S., Leshin, L.~A., 2006. {Meteorites and the
  Timing, Mechanisms, and Conditions of Terrestrial Planet Accretion and Early
  Differentiation}. Meteorites and the Early Solar System II, 53--66.

\bibitem[{Michel(1997)}]{1997-Icarus-129-Michel}
Michel, P., 1997. Effects of linear secular resonances in the region of
  semimajor axes smaller than 2 \{AU\}. Icarus 129~(2), 348 -- 366.

\bibitem[{{Michel} et~al.(2000){Michel}, {Migliorini}, {Morbidelli}, and
  {Zappal{\`a}}}]{2000-Icarus-145-Michel}
{Michel}, P., {Migliorini}, F., {Morbidelli}, A., {Zappal{\`a}}, V., Jun. 2000.
  {The Population of Mars-Crossers: Classification and Dynamical Evolution}.
  Icarus 145, 332--347.

\bibitem[{{Milani} et~al.(2000){Milani}, {Chesley}, and
  {Valsecchi}}]{2000-PSS-48-Milani}
{Milani}, A., {Chesley}, S.~R., {Valsecchi}, G.~B., Aug. 2000. {Asteroid close
  encounters with Earth: risk assessment}. Planetary and Space Science 48,
  945--954.

\bibitem[{Milani et~al.(2008)Milani, Gronchi, Farnocchia, Kne\v{z}evi{\'c},
  Jedicke, Denneau, and Pierfederici}]{milani2008topocentric}
Milani, A., Gronchi, G., Farnocchia, D., Kne\v{z}evi{\'c}, Z., Jedicke, R.,
  Denneau, L., Pierfederici, F., 2008. Topocentric orbit determination:
  algorithms for the next generation surveys. Icarus 195~(1), 474--492.

\bibitem[{{Morbidelli} et~al.(2002){Morbidelli}, {Bottke}, {Froeschl{\'e}}, and
  {Michel}}]{2002-AsteroidsIII-4-Morbidelli}
{Morbidelli}, A., {Bottke}, Jr., W.~F., {Froeschl{\'e}}, C., {Michel}, P.,
  2002. {Origin and Evolution of Near-Earth Objects}. Asteroids III, 409--422.

\bibitem[{{Morbidelli} and {Nesvorn{\'y}}(1999)}]{1999-Icarus-139-Morbidelli}
{Morbidelli}, A., {Nesvorn{\'y}}, D., Jun. 1999. {Numerous Weak Resonances
  Drive Asteroids toward Terrestrial Planets Orbits}. Icarus 139, 295--308.

\bibitem[{Mueller et~al.(2011)Mueller, Delbo, Hora, Trilling, Bhattacharya,
  Bottke, Chesley, Emery, Fazio, Harris, Mainzer, Mommert, Penprase, Smith,
  Spahr, Stansberry, and Thomas}]{2011-AJ-141-Mueller}
Mueller, M., Delbo, M., Hora, J.~L., Trilling, D.~E., Bhattacharya, B., Bottke,
  W.~F., Chesley, S., Emery, J.~P., Fazio, G., Harris, A.~W., Mainzer, A.,
  Mommert, M., Penprase, B., Smith, H.~A., Spahr, T.~B., Stansberry, J.~A.,
  Thomas, C.~A., Apr. 2011. {ExploreNEOs. III. Physical Characterization of 65
  Potential Spacecraft Target Asteroids}. Astronomical Journal 141, 109.

\bibitem[{Murdoch et~al.(2012)Murdoch, Abell, Carnelli, Carry, Cheng,
  Drolshagen, Fontaine, Galvez, Koschny, Kueppers, Michel, Reed, and
  Ulamec}]{2012-ESA-AIDA}
Murdoch, N., Abell, P., Carnelli, I., Carry, B., Cheng, A., Drolshagen, G.,
  Fontaine, M., Galvez, A., Koschny, D., Kueppers, M., Michel, P., Reed, C.,
  Ulamec, S., 2012. Asteroid impact \& deflection assessment (aida) mission.
  Tech. rep., European Space Agency.

\bibitem[{{Nakamura} et~al.(2011){Nakamura}, {Noguchi}, {Tanaka}, {Zolensky},
  {Kimura}, {Tsuchiyama}, {Nakato}, {Ogami}, {Ishida}, {Uesugi}, {Yada},
  {Shirai}, {Fujimura}, {Okazaki}, {Sandford}, {Ishibashi}, {Abe}, {Okada},
  {Ueno}, {Mukai}, {Yoshikawa}, and {Kawaguchi}}]{2011-Science-333-Nakamura}
{Nakamura}, T., {Noguchi}, T., {Tanaka}, M., {Zolensky}, M.~E., {Kimura}, M.,
  {Tsuchiyama}, A., {Nakato}, A., {Ogami}, T., {Ishida}, H., {Uesugi}, M.,
  {Yada}, T., {Shirai}, K., {Fujimura}, A., {Okazaki}, R., {Sandford}, S.~A.,
  {Ishibashi}, Y., {Abe}, M., {Okada}, T., {Ueno}, M., {Mukai}, T.,
  {Yoshikawa}, M., {Kawaguchi}, J., Aug 2011. {Itokawa Dust Particles: A Direct
  Link Between S-Type Asteroids and Ordinary Chondrites}. Science 333,
  1113--1115.

\bibitem[{Nesvorny(2012)}]{PDS-Nesvorny}
Nesvorny, D., 2012. Nesvorny hcm asteroid families v2.0. NASA Planetary Data
  System, eAR-A-VARGBDET-5-NESVORNYFAM-V2.0.

\bibitem[{{Nesvorn{\'y}} et~al.(2010){Nesvorn{\'y}}, {Bottke},
  {Vokrouhlick{\'y}}, {Chapman}, and {Rafkin}}]{2010-Icarus-209-Nesvorny}
{Nesvorn{\'y}}, D., {Bottke}, W.~F., {Vokrouhlick{\'y}}, D., {Chapman}, C.~R.,
  {Rafkin}, S., Oct. 2010. {Do planetary encounters reset surfaces of near
  Earth asteroids?} Icarus 209, 510--519.

\bibitem[{{Nesvorn{\'y}} et~al.(2005){Nesvorn{\'y}}, {Jedicke}, {Whiteley}, and
  {Ivezi{\'c}}}]{2005-Icarus-173-Nesvorny}
{Nesvorn{\'y}}, D., {Jedicke}, R., {Whiteley}, R.~J., {Ivezi{\'c}}, {\v Z}.,
  Jan. 2005. {Evidence for asteroid space weathering from the Sloan Digital Sky
  Survey}. Icarus 173, 132--152.

\bibitem[{Ostro et~al.(2002)Ostro, Hudson, Benner, Giorgini, Magri, Margot, and
  Nolan}]{2002-AsteroidsIII-2.2-Ostro}
Ostro, S.~J., Hudson, R.~S., Benner, L.~A.~M., Giorgini, J.~D., Magri, C.,
  Margot, J.~L., Nolan, M.~C., 2002. {Asteroid Radar Astronomy}. Asteroids III,
  151--168.

\bibitem[{Parker et~al.(2008)Parker, Ivezi{\'c}, Juri{\'c}, Lupton, Sekora, and
  Kowalski}]{2008-Icarus-198-Parker}
Parker, A., Ivezi{\'c}, {\v Z}., Juri{\'c}, M., Lupton, R., Sekora, M.~D.,
  Kowalski, A., Nov 2008. {The size distributions of asteroid families in the
  SDSS Moving Object Catalog 4}. Icarus 198, 138--155.

\bibitem[{Perko et~al.(2001)Perko, Nelson, and Sadeh}]{perko2001surface}
Perko, H.~A., Nelson, J.~D., Sadeh, W.~Z., 2001. Surface cleanliness effect on
  lunar soil shear strength. Journal of geotechnical and geoenvironmental
  engineering 127~(4), 371--383.

\bibitem[{{Polishook} et~al.(2012){Polishook}, {Binzel}, {Lockhart}, {DeMeo},
  {Golisch}, {Bus}, and {Gulbis}}]{2012-Icarus-221-Polishook}
{Polishook}, D., {Binzel}, R.~P., {Lockhart}, M., {DeMeo}, F.~E., {Golisch},
  W., {Bus}, S.~J., {Gulbis}, A.~A.~S., Nov. 2012. {Spectral and spin
  measurement of two small and fast-rotating near-Earth asteroids}. Icarus 221,
  1187--1189.

\bibitem[{{Popescu} et~al.(2011){Popescu}, {Birlan}, {Binzel}, {Vernazza},
  {Barucci}, {Nedelcu}, {DeMeo}, and {Fulchignoni}}]{2011-AA-535-Popescu}
{Popescu}, M., {Birlan}, M., {Binzel}, R., {Vernazza}, P., {Barucci}, A.,
  {Nedelcu}, D.~A., {DeMeo}, F., {Fulchignoni}, M., Nov. 2011. {Spectral
  properties of eight near-Earth asteroids}. Astronomy and Astrophysics 535,
  A15.

\bibitem[{Pravec et~al.(2006)Pravec, Scheirich, Ku{\v s}nir{\'a}k, {\v
  S}arounov{\'a}, Mottola, Hahn, Brown, Esquerdo, Kaiser, Krzeminski, Pray,
  Warner, Harris, Nolan, Howell, Benner, Margot, Gal{\'a}d, Holliday, Hicks,
  Krugly, Tholen, Whiteley, Marchis, Degraff, Grauer, Larson, Velichko, Cooney,
  Stephens, Zhu, Kirsch, Dyvig, Snyder, Reddy, Moore, Gajdo{\v s}, Vil{\'a}gi,
  Masi, Higgins, Funkhouser, Knight, Slivan, Behrend, Grenon, Burki, Roy,
  Demeautis, Matter, Waelchli, Revaz, Klotz, Rieugn{\'e}, Thierry, Cotrez,
  Brunetto, and Kober}]{2006-Icarus-181-Pravec}
Pravec, P., Scheirich, P., Ku{\v s}nir{\'a}k, P., {\v S}arounov{\'a}, L.,
  Mottola, S., Hahn, G., Brown, P.~G., Esquerdo, G.~A., Kaiser, N., Krzeminski,
  Z., Pray, D.~P., Warner, B.~D., Harris, A.~W., Nolan, M.~C., Howell, E.~S.,
  Benner, L.~A.~M., Margot, J.-L., Gal{\'a}d, A., Holliday, W., Hicks, M.~D.,
  Krugly, Y.~N., Tholen, D.~J., Whiteley, R.~J., Marchis, F., Degraff, D.~R.,
  Grauer, A., Larson, S., Velichko, F.~P., Cooney, W.~R., Stephens, R., Zhu,
  J., Kirsch, K., Dyvig, R., Snyder, L., Reddy, V., Moore, S., Gajdo{\v s}, {\v
  S}., Vil{\'a}gi, J., Masi, G., Higgins, D., Funkhouser, G., Knight, B.,
  Slivan, S.~M., Behrend, R., Grenon, M., Burki, G., Roy, R., Demeautis, C.,
  Matter, D., Waelchli, N., Revaz, Y., Klotz, A., Rieugn{\'e}, M., Thierry, P.,
  Cotrez, V., Brunetto, L., Kober, G., Mar 2006. {Photometric survey of binary
  near-Earth asteroids}. Icarus 181, 63--93.

\bibitem[{Reddy et~al.(2015)Reddy, Dunn, Thomas, Moskovitz, and
  Burbine}]{2015-AsteroidsIV-Reddy}
Reddy, V., Dunn, T., Thomas, C., Moskovitz, N., Burbine, T., 2015. {Mineralogy
  and Surface Composition of Asteroids}. Asteroids IV, na.

\bibitem[{{Reddy} et~al.(2011){Reddy}, {Nathues}, and
  {Gaffey}}]{2011-Icarus-212-Reddy}
{Reddy}, V., {Nathues}, A., {Gaffey}, M.~J., Mar. 2011. {First fragment of
  Asteroid 4 Vesta's mantle detected}. Icarus 212, 175--179.

\bibitem[{{Ribeiro} et~al.(2014){Ribeiro}, {Roig}, {Ca{\~n}ada-Assandri},
  {Carvano}, {Jasmin}, {Alvarez-Candal}, and
  {Gil-Hutton}}]{2014-PSS-92-Ribeiro}
{Ribeiro}, A.~O., {Roig}, F., {Ca{\~n}ada-Assandri}, M., {Carvano}, J.~M.~F.,
  {Jasmin}, F.~L., {Alvarez-Candal}, A., {Gil-Hutton}, R., Mar. 2014. {The
  first confirmation of V-type asteroids among the Mars crosser population}.
  Planetary and Space Science 92, 57--64.

\bibitem[{{Rivkin} et~al.(2011){Rivkin}, {Thomas}, {Trilling}, {Enga}, and
  {Grier}}]{2011-Icarus-211-Rivkin}
{Rivkin}, A.~S., {Thomas}, C.~A., {Trilling}, D.~E., {Enga}, M.-t., {Grier},
  J.~A., Feb. 2011. {Ordinary chondrite-like colors in small Koronis family
  members}. Icarus 211, 1294--1297.

\bibitem[{{Roig} et~al.(2008){Roig}, {Ribeiro}, and
  {Gil-Hutton}}]{2008-AA-483-Roig}
{Roig}, F., {Ribeiro}, A.~O., {Gil-Hutton}, R., Jun. 2008. {Taxonomy of
  asteroid families among the Jupiter Trojans: comparison between spectroscopic
  data and the Sloan Digital Sky Survey colors}. Astronomy and Astrophysics
  483, 911--931.

\bibitem[{{Rudawska} et~al.(2012){Rudawska}, {Vaubaillon}, and
  {Atreya}}]{2012-AA-541-Rudawska}
{Rudawska}, R., {Vaubaillon}, J., {Atreya}, P., May 2012. {Association of
  individual meteors with their parent bodies}. Astronomy and Astrophysics 541,
  A2.

\bibitem[{{Sanchez} et~al.(2013){Sanchez}, {Michelsen}, {Reddy}, and
  {Nathues}}]{2013-Icarus-225-Sanchez}
{Sanchez}, J.~A., {Michelsen}, R., {Reddy}, V., {Nathues}, A., Jul. 2013.
  {Surface composition and taxonomic classification of a group of near-Earth
  and Mars-crossing asteroids}. Icarus 225, 131--140.

\bibitem[{{Sanchez} et~al.(2012){Sanchez}, {Reddy}, {Nathues}, {Cloutis},
  {Mann}, and {Hiesinger}}]{2012-Icarus-220-Sanchez}
{Sanchez}, J.~A., {Reddy}, V., {Nathues}, A., {Cloutis}, E.~A., {Mann}, P.,
  {Hiesinger}, H., Jul. 2012. {Phase reddening on near-Earth asteroids:
  Implications for mineralogical analysis, space weathering and taxonomic
  classification}. Icarus 220, 36--50.

\bibitem[{{S{\'a}nchez} and {Scheeres}(2014)}]{2014-MPS-49-Sanchez}
{S{\'a}nchez}, P., {Scheeres}, D.~J., May 2014. {The strength of regolith and
  rubble pile asteroids}. Meteoritics and Planetary Science 49, 788--811.

\bibitem[{Sasaki et~al.(2001)Sasaki, Nakamura, Hamabe, Kurahashi, and
  Hiroi}]{2001-Nature-410-Sasaki}
Sasaki, S., Nakamura, K., Hamabe, Y., Kurahashi, E., Hiroi, T., Mar. 2001.
  {Production of iron nanoparticles by laser irradiation in a simulation of
  lunar-like space weathering}. Nature 410, 555--557.

\bibitem[{{Scheeres} et~al.(2010){Scheeres}, {Hartzell}, {S{\'a}nchez}, and
  {Swift}}]{2010-Icarus-210-Scheeres}
{Scheeres}, D.~J., {Hartzell}, C.~M., {S{\'a}nchez}, P., {Swift}, M., Dec.
  2010. {Scaling forces to asteroid surfaces: The role of cohesion}. Icarus
  210, 968--984.

\bibitem[{{Sharma} et~al.(2006){Sharma}, {Jenkins}, and
  {Burns}}]{2006-Icarus-183-Sharma}
{Sharma}, I., {Jenkins}, J.~T., {Burns}, J.~A., Aug. 2006. {Tidal encounters of
  ellipsoidal granular asteroids with planets}. Icarus 183, 312--330.

\bibitem[{{Solano} et~al.(2013){Solano}, {Rodrigo}, {Pulido}, and
  {Carry}}]{2013-AN--Solano}
{Solano}, E., {Rodrigo}, C., {Pulido}, R., {Carry}, B., Feb 2013. {Precovery of
  near-Earth asteroids by a citizen-science project of the Spanish Virtual
  Observatory}. Astronomische Nachrichten 335, 142--149.

\bibitem[{Spoto et~al.(2014)Spoto, Milani, Farnocchia, Chesley, Micheli,
  Valsecchi, Perna, and Hainaut}]{2014-AA-572-Spoto}
Spoto, F., Milani, A., Farnocchia, D., Chesley, S.~R., Micheli, M., Valsecchi,
  G.~B., Perna, D., Hainaut, O., 2014. Nongravitational perturbations and
  virtual impactors: the case of asteroid (410777) 2009 fd. Astronomy \&
  Astrophysics 572, A100.

\bibitem[{{Stokes} et~al.(2000){Stokes}, {Evans}, {Viggh}, {Shelly}, and
  {Pearce}}]{2000-Icarus-148-Stokes}
{Stokes}, G.~H., {Evans}, J.~B., {Viggh}, H.~E.~M., {Shelly}, F.~C., {Pearce},
  E.~C., Nov 2000. {Lincoln Near-Earth Asteroid Program (LINEAR)}. Icarus 148,
  21--28.

\bibitem[{Strazzulla et~al.(2005)Strazzulla, Dotto, Binzel, Brunetto, Barucci,
  Blanco, and Orofino}]{2005-Icarus-174-Strazzulla}
Strazzulla, G., Dotto, E., Binzel, R.~P., Brunetto, R., Barucci, M.~A., Blanco,
  A., Orofino, V., mar 2005. {Spectral alteration of the Meteorite Epinal (H5)
  induced by heavy ion irradiation: a simulation of space weathering effects on
  near-Earth asteroids}. Icarus 174, 31--35.

\bibitem[{{Tholen}(1984)}]{1984-PhD-Tholen}
{Tholen}, D.~J., 1984. {Asteroid taxonomy from cluster analysis of Photometry}.
  Ph.D. thesis, Arizona Univ., Tucson.

\bibitem[{Tholen and Barucci(1989)}]{1989-AsteroidsII-Tholen}
Tholen, D.~J., Barucci, M.~A., 1989. {Asteroid taxonomy}. Asteroids II,
  298--315.

\bibitem[{{Thomas} and {Binzel}(2010)}]{2010-Icarus-205-Thomas}
{Thomas}, C.~A., {Binzel}, R.~P., Feb. 2010. {Identifying meteorite source
  regions through near-Earth object spectroscopy}. Icarus 205, 419--429.

\bibitem[{{Thomas} et~al.(2011){Thomas}, {Rivkin}, {Trilling}, {Marie-therese
  Enga}, and {Grier}}]{2011-Icarus-212-Thomas}
{Thomas}, C.~A., {Rivkin}, A.~S., {Trilling}, D.~E., {Marie-therese Enga},
  {Grier}, J.~A., Mar. 2011. {Space weathering of small Koronis family
  members}. Icarus 212, 158--166.

\bibitem[{{Thomas} et~al.(2012){Thomas}, {Trilling}, and
  {Rivkin}}]{2012-Icarus-219-Thomas}
{Thomas}, C.~A., {Trilling}, D.~E., {Rivkin}, A.~S., May 2012. {Space
  weathering of small Koronis family asteroids in the SDSS Moving Object
  Catalog}. Icarus 219, 505--507.

\bibitem[{Vernazza et~al.(2009)Vernazza, Binzel, Rossi, Fulchignoni, and
  Birlan}]{2009-Nature-458-Vernazza}
Vernazza, P., Binzel, R.~P., Rossi, A., Fulchignoni, M., Birlan, M., Apr. 2009.
  {Solar wind as the origin of rapid reddening of asteroid surfaces}. Nature
  458, 993--995.

\bibitem[{Vernazza et~al.(2008)Vernazza, Binzel, Thomas, DeMeo, Bus, Rivkin,
  and Tokunaga}]{2008-Nature-454-Vernazza}
Vernazza, P., Binzel, R.~P., Thomas, C.~A., DeMeo, F.~E., Bus, S.~J., Rivkin,
  A.~S., Tokunaga, A.~T., Aug. 2008. {Compositional differences between
  meteorites and near-Earth asteroids}. Nature 454, 858--860.

\bibitem[{{Vernazza} et~al.(2012){Vernazza}, {Delbo}, {King}, {Izawa},
  {Olofsson}, {Lamy}, {Cipriani}, {Binzel}, {Marchis}, {Mer{\'{\i}}n}, and
  {Tamanai}}]{2012-Icarus-221-Vernazza}
{Vernazza}, P., {Delbo}, M., {King}, P.~L., {Izawa}, M.~R.~M., {Olofsson}, J.,
  {Lamy}, P., {Cipriani}, F., {Binzel}, R.~P., {Marchis}, F., {Mer{\'{\i}}n},
  B., {Tamanai}, A., Nov. 2012. {High surface porosity as the origin of
  emissivity features in asteroid spectra}. Icarus 221, 1162--1172.

\bibitem[{Veverka et~al.(2000)Veverka, Robinson, Thomas, Murchie, Bell,
  Izenberg, Chapman, Harch, Bell, Carcich, Cheng, Clark, Domingue, Dunham,
  Farquhar, Gaffey, Hawkins, Joseph, Kirk, Li, Lucey, Malin, Martin, McFadden,
  Merline, Miller, Owen, Peterson, Prockter, Warren, Wellnitz, Williams, and
  Yeomans}]{2000-Science-289-Veverka}
Veverka, J., Robinson, M., Thomas, P., Murchie, S., Bell, J.~F., Izenberg, N.,
  Chapman, C., Harch, A., Bell, M., Carcich, B., Cheng, A., Clark, B.,
  Domingue, D., Dunham, D., Farquhar, R., Gaffey, M.~J., Hawkins, E., Joseph,
  J., Kirk, R., Li, H., Lucey, P., Malin, M., Martin, P., McFadden, L.,
  Merline, W.~J., Miller, J.~K., Owen, W.~M., Peterson, C., Prockter, L.,
  Warren, J., Wellnitz, D., Williams, B.~G., Yeomans, D.~K., Sep. 2000. {NEAR
  at Eros: Imaging and Spectral Results}. Science 289, 2088--2097.

\bibitem[{{Wetherill}(1979)}]{1979-Icarus-37-Wetherill}
{Wetherill}, G.~W., Jan. 1979. {Steady state populations of Apollo-Amor
  objects}. Icarus 37, 96--112.

\bibitem[{{Willman} and {Jedicke}(2011)}]{2011-Icarus-211-Willman}
{Willman}, M., {Jedicke}, R., Jan. 2011. {Asteroid age distributions determined
  by space weathering and collisional evolution models}. Icarus 211, 504--510.

\bibitem[{{Willman} et~al.(2010){Willman}, {Jedicke}, {Moskovitz},
  {Nesvorn{\'y}}, {Vokrouhlick{\'y}}, and
  {Moth{\'e}-Diniz}}]{2010-Icarus-208-Willman}
{Willman}, M., {Jedicke}, R., {Moskovitz}, N., {Nesvorn{\'y}}, D.,
  {Vokrouhlick{\'y}}, D., {Moth{\'e}-Diniz}, T., Aug. 2010. {Using the youngest
  asteroid clusters to constrain the space weathering and gardening rate on
  S-complex asteroids}. Icarus 208, 758--772.

\bibitem[{{Willman} et~al.(2008){Willman}, {Jedicke}, {Nesvorn{\'y}},
  {Moskovitz}, {Ivezi{\'c}}, and {Fevig}}]{2008-Icarus-195-Willman}
{Willman}, M., {Jedicke}, R., {Nesvorn{\'y}}, D., {Moskovitz}, N.,
  {Ivezi{\'c}}, {\v Z}., {Fevig}, R., Jun. 2008. {Redetermination of the space
  weathering rate using spectra of Iannini asteroid family members}. Icarus
  195, 663--673.

\bibitem[{{Wisdom}(1983)}]{1983-Icarus-56-Wisdom}
{Wisdom}, J., Oct. 1983. {Chaotic behavior and the origin of the 3/1 Kirkwood
  gap}. Icarus 56, 51--74.

\bibitem[{{Yang} and {Jewitt}(2007)}]{2007-AJ-134-Yang}
{Yang}, B., {Jewitt}, D., Jul. 2007. {Spectroscopic Search for Water Ice on
  Jovian Trojan Asteroids}. Astronomical Journal 134, 223--228.

\bibitem[{{Yang} and {Jewitt}(2011)}]{2011-AJ-141-Yang}
{Yang}, B., {Jewitt}, D., Mar. 2011. {A Near-infrared Search for Silicates in
  Jovian Trojan Asteroids}. Astronomical Journal 141, 95.

\bibitem[{{Yano} et~al.(2010){Yano}, {Yoshikawa}, {Yano}, {Tsuda}, {Nakazawa},
  {Mimamino}, {Terui}, {Saiki}, {Nishiyama}, {Kubota}, {Okada}, {Morimoto},
  {Ogawa}, {Okamoto}, {Takagi}, {Tachibana}, {Nakamura}, {Hirata}, and
  {Demura}}]{2010-COSPAR-38-Yano}
{Yano}, H., {Yoshikawa}, M., {Yano}, H., {Tsuda}, Y., {Nakazawa}, S.,
  {Mimamino}, H., {Terui}, F., {Saiki}, T., {Nishiyama}, K., {Kubota}, T.,
  {Okada}, T., {Morimoto}, M.~Y., {Ogawa}, N., {Okamoto}, C., {Takagi}, Y.,
  {Tachibana}, S., {Nakamura}, R., {Hirata}, N., {Demura}, H., 2010.
  {Hayabusa's follow-on mission for surface and sub-surface sample return from
  a C-type NEO}. In: 38th COSPAR Scientific Assembly. Vol.~38 of COSPAR
  Meeting. p. 635.

\bibitem[{{Ye}(2011)}]{2011-AJ-141-Ye}
{Ye}, Q.-z., Feb. 2011. {BVRI Photometry of 53 Unusual Asteroids}. Astronomical
  Journal 141, 32.

\bibitem[{Yoshida(1990)}]{yoshida1990construction}
Yoshida, H., 1990. Construction of higher order symplectic integrators. Physics
  Letters A 150~(5), 262--268.

\bibitem[{{Yu} et~al.(2014){Yu}, {Richardson}, {Michel}, {Schwartz}, and
  {Ballouz}}]{2014-Icarus-242-Yu}
{Yu}, Y., {Richardson}, D.~C., {Michel}, P., {Schwartz}, S.~R., {Ballouz},
  R.-L., Nov. 2014. {Numerical predictions of surface effects during the 2029
  close approach of Asteroid 99942 Apophis}. Icarus 242, 82--96.

\end{thebibliography}


\appendix

\section{Forces model\label{app:forces}}

  \indent In the following we will briefly discuss the force models
  used in equations (\ref{eq:fbaleasy}) and (\ref{eq:fbalhard}). 

  \subsection{Self gravity of the asteroid}
    \indent On the surface of the asteroid a particle of mass  $\mu$
    will experience the following gravitational pull towards the
    asteroid's center of mass: 
    \begin{equation}
      F_{ga}= \mathcal{G} \mu m / R^2 \label{eq:grav}
    \end{equation}
    \noindent where $\mathcal{G}$ is the gravitational constant, $m$
    the mass and $R$ the local radius of the asteroid.  

  \subsection{Forces due to the Rotating Frame}
    \indent In a rotating frame of reference, surface particles are
    subject to a centrifugal force $F_{cf}$. 
    It will suffice to consider principle axis rotation as we are
    primarily interested in best and worst case scenarios.  
    Therefore, the particle feeling the maximum centrifugal
    acceleration then is located farthest from the axis of rotation
    and we have   
    \begin{equation}
      F_{cf}= \mu R \omega^2 \cos{\phi} \label{eq:rot}
    \end{equation} 
    \noindent Here, $\omega$ is the spin rate and $\phi$ the angle
    between the asteroid's spin axis and the vector $\vec r$
    connecting the centers of gravity of the planet and the
    asteroid. In other words $\phi$ is the asteroid’s obliquity with
    respect to is planetocentric orbit. 
    Consequently, $F_{cf}^{max}=\mu R^{max} \omega_{sb}^2$. This
    corresponds to a particle on the equator of a spherical asteroid
    with maximum radius $R^{max}$ and a rotation rate close to the
    spin barrier $\omega_{sb}$. 
    As we assume that the particles are at rest on the surface, no
    Coriolis forces will have to be accounted for. 

  \subsection{Tidal Acceleration}
    \indent Asteroid that have close encounters with other massive
    bodies will experience tidal forces.  
    Towards an asteroid's surface the tidal acceleration becomes
    stronger compared to volume elements that are closer to the center
    of gravity. 
    Consequently surface particles are pulled away more easily than
    say boulders close to the center of the asteroid. A simple model for
    the tidal force acting on a particle with mass $\mu$ is:
    \begin{equation}
      F_{td}=2R \mathcal{G} \mu M / D^3 \label{eq:tide}
    \end{equation}
    \noindent where $M$ is the mass of the planet and $D$ the distance
    between the asteroid and the planet. In principle, there are
    additional tidal contributions from the sun, and the moon if the
    Earth is approached. 
    Those are, however, small compared to the contribution of the
    Earth itself and shall, therefore, be neglected.

  \subsection{Particle Cohesion\label{app:cohesion}}
    \indent The role of cohesion in rubble pile asteroids has been
    studied extensively over the past years
    \citep{2010-Icarus-210-Scheeres,2011-PSS-59-Hartzell,2014-MPS-49-Sanchez}. 
    It is currently understood that particle cohesion plays and important
    role in determining limits for rubble-pile asteroid failures,
    fission, 
    and for the survival of contact binaries.   
    The standard cohesion model as presented in
    \cite{2011-PSS-59-Hartzell} is based on van der Waals interaction in
    granular materials.  
    The inter particle force can be modeled as: 
    \begin{equation}
      F_{co}=C S^2 \sigma \label{eq:cohesion}
    \end{equation}
    \noindent where $C=5.14 \times 10^{-2}\; kg/s^2$ is a material
    constant related to the Hamaker constant, $S\sim1$ is the
    so-called cleanliness ratio, a qualitative  
    factor indicating the impurity of granular surface coatings and
    $\sigma$ is the radius of the particles on the asteroid's
    surface. For a detailed  
    discussion of equation \ref{eq:cohesion} see
    e.g. \cite{2011-PSS-59-Hartzell}. 
    As it is harder to free larger particles or rocks from the
    surface, we will increase the particle radius that shall be lifted
    against cohesion. 
    Equation (\ref{eq:cohesion}) illustrates that the larger the
    particles and the cleaner the surface, the larger the cohesive
    force. Let us now determine the 
    approximate size of the largest particles that will be relevant
    for our resurfacing considerations. For this purpose we assume an
    $N\propto (2\sigma)^{-3}$ distribution of the  
    number of particles $N$ on the asteroid's surface.
    Using equation (36) in \citet{2014-MPS-49-Sanchez} we can calculate
    the particle diameter up to which smaller particles cover 99\% of
    the surface. 
    This is $\sigma=100 r_0$, where we furthermore assume that $r_0$ is of the order of the 
    smallest particle radii that could be detected on Itokawa
    ($r_0\simeq10^{-6}$ m). Hence, practically all of the asteroid's
    surface is covered by particles 
    smaller than 0.1 millimeters.

\pagebreak
\onecolumn
\section{Taxonomic classifications\label{app:taxo}}

\setcounter{table}{0}

\begin{center}
    \tablefirsthead{\hline
    \hline
    \multicolumn{1}{c}{\#} &
    \multicolumn{1}{c}{Name} & Class &
    \r & $\delta$\r &
    \i & $\delta$\i &
    \z & $\delta$\z &
    \multicolumn{1}{c}{a} &
    \multicolumn{1}{c}{e} &
    \multicolumn{1}{c}{i} &
    H \\
    &&&&&&&&& (au) & (\degr)\\
    \hline}

    \tablehead{\hline
    \hline
    \multicolumn{1}{c}{\#} &
    \multicolumn{1}{c}{Name} & Class &
    \r & $\delta$\r &
    \i & $\delta$\i &
    \z & $\delta$\z &
    \multicolumn{1}{c}{a} &
    \multicolumn{1}{c}{e} &
    \multicolumn{1}{c}{i} &
    H \\
    \hline}

    \tabletail{%
      \hline
      \multicolumn{13}{r}{\small\sl continued on next page}\\
      \hline}

    \tablelasttail{\hline}
    \bottomcaption{
    The 982 asteroids with four-bands photometry classified in this
    work. For each, we list its IAU number and designation, taxonomic
    class, \r, \i, and \z\ filter reflectance, normalized to \g, the
    orbital elements (semi-major axis a, eccentricity e, inclination
    i), and the absolute magnitude.
    \label{tab: class4}}

\begin{supertabular}{rlc cc cc cc rrr c}
     58070 & 1034 T-2        & C   &    1.029 &    0.065 &    1.199 &    0.102 &    1.127 &    0.148 & 2.347 & 0.301 &   6.14 & 16.00 \\ 
 30717 & 1937 UD         & S   &    1.282 &    0.067 &    1.424 &    0.048 &    1.192 &    0.058 & 2.168 & 0.334 &   8.96 & 14.00 \\ 
118162 & 1951 SX         & S   &    1.200 &    0.042 &    1.363 &    0.069 &    1.308 &    0.106 & 1.875 & 0.254 &  39.11 & 15.00 \\ 
216464 & 1974 PB         & V   &    1.237 &    0.038 &    1.416 &    0.056 &    1.104 &    0.071 & 2.318 & 0.341 &   6.57 & 17.00 \\ 
136564 & 1977 VA         & C   &    1.028 &    0.038 &    1.107 &    0.051 &    1.096 &    0.071 & 1.866 & 0.394 &   2.98 & 19.20 \\ 
159368 & 1979 QB         & S   &    1.207 &    0.023 &    1.272 &    0.002 &    1.085 &    0.031 & 2.330 & 0.443 &   3.35 & 17.00 \\ 
120450 & 1982 SV         & S   &    1.225 &    0.034 &    1.343 &    0.062 &    1.236 &    0.102 & 2.418 & 0.367 &  21.86 & 15.80 \\ 
  9292 & 1982 UE2        & S   &    1.203 &    0.006 &    1.345 &    0.008 &    1.205 &    0.003 & 2.196 & 0.285 &   4.34 & 15.00 \\
\end{supertabular}
\end{center}

\begin{center}
    \tablefirsthead{\hline
    \hline
    \multicolumn{1}{c}{\#} &
    \multicolumn{1}{c}{Name} & Class &
    \r & $\delta$\r &
    \i & $\delta$\i &
    \multicolumn{1}{c}{a} &
    \multicolumn{1}{c}{e} &
    \multicolumn{1}{c}{i} &
    H \\
    &&&&&&& (au) & (\degr)\\
    \hline}

    \tablehead{\hline
    \hline
    \multicolumn{1}{c}{\#} &
    \multicolumn{1}{c}{Name} & Class &
    \r & $\delta$\r &
    \i & $\delta$\i &
    \multicolumn{1}{c}{a} &
    \multicolumn{1}{c}{e} &
    \multicolumn{1}{c}{i} &
    H \\
    \hline}

    \tabletail{%
      \hline
      \multicolumn{11}{r}{\small\sl continued on next page}\\
      \hline}

    \tablelasttail{\hline}
    \bottomcaption{
    The 254 asteroids with three-bands photometry classified in this
    work. For each, we list its IAU number and designation, taxonomic
    class, \r\ and \i~filter reflectance, normalized to \g, the
    orbital elements (semi-major axis a, eccentricity e, inclination
    i), and the absolute magnitude.
    \label{tab: class3}}

\begin{supertabular}{rlc cc cc rrr c}
     30717 & 1937 UD         & U   &    1.215 &    0.048 &    1.330 &    0.075 & 2.168 & 0.334 &   8.96 & 14.00 \\ 
 73673 & 1986 RX1        & S   &    1.170 &    0.034 &    1.277 &    0.050 & 2.553 & 0.363 &   7.04 & 14.00 \\ 
100017 & 1989 TN2        & C   &    0.958 &    0.030 &    1.014 &    0.048 & 2.306 & 0.295 &   8.06 & 16.00 \\ 
 69274 & 1989 UZ1        & V   &    1.153 &    0.037 &    1.311 &    0.057 & 2.278 & 0.319 &   9.06 & 15.00 \\ 
  5828 & 1991 AM         & V   &    1.125 &    0.027 &    1.329 &    0.046 & 1.697 & 0.696 &  30.11 & 15.00 \\ 
 85274 & 1994 GH         & A   &    1.401 &    0.041 &    1.661 &    0.067 & 2.149 & 0.247 &   5.80 & 17.00 \\ 
  9881 & 1994 SE         & V   &    1.243 &    0.049 &    1.162 &    0.062 & 2.342 & 0.410 &   8.08 & 15.00 \\ 
     0 & 1997 CO5        & S   &    1.300 &    0.033 &    1.502 &    0.055 & 2.617 & 0.485 &  19.08 & 16.00 \\
\end{supertabular}
\end{center}

\end{document}